\documentclass[10pt]{article}

\usepackage[preprint]{tmlr}

\usepackage{amsmath,amsfonts,bm}

\def\eqref#1{equation~\ref{#1}}

\def\1{\bm{1}}

\DeclareMathAlphabet{\mathsfit}{\encodingdefault}{\sfdefault}{m}{sl}
\SetMathAlphabet{\mathsfit}{bold}{\encodingdefault}{\sfdefault}{bx}{n}

\usepackage{hyperref}
\usepackage{wrapfig}
\usepackage{url}
\usepackage{graphicx}
\usepackage{wrapfig}
\usepackage{hyperref}
\usepackage{subcaption}
\usepackage{booktabs}
\usepackage{multirow}
\usepackage{amssymb}
\usepackage{xcolor}
\usepackage{url}
\colorlet{green}{green!60!black}

\title{Airfoil2Vec: Spectral Geometry-Conditioned \\ Neural Surrogate Models for Airfoil Aerodynamics \\ and a Downforce-Generating CFD Dataset}

\author{\name Haitz Sáez de Ocáriz Borde, Flavio Savarino, Andrei Cristian Popescu, Pietro Innocenzi, Pantelis Papageorgiou, Xerxes Xian Chong \\
      \addr Ratio Labs
}

\def\month{MM}  
\def\year{YYYY} 
\def\openreview{\url{https://openreview.net/forum?id=XXXX}} 

\begin{document}
\maketitle

\begin{abstract}
We introduce a dataset of approximately 10,000 Reynolds-Averaged Navier--Stokes (RANS) simulations of steady, incompressible, two-dimensional subsonic flow around downforce-generating NACA 4-digit airfoils, targeting aerodynamic regimes relevant to automotive and motorsport applications (openly available on \url{https://huggingface.co/datasets/ratiolabs/downforce-airfoils}). Using this resource, we study geometry-conditioned neural surrogates for fast flow prediction, comparing neural fields with neural ODEs, MLPs with graph-based models, and several spectral geometry-conditioning methods. We further propose Airfoil2Vec, an airfoil-specific spectral geometry encoder that combines the joint contour spectrum with separate spectral representations of camber and thickness, for predicting continuous pressure and velocity fields. We evaluate generalization through angle-of-attack interpolation, interpolation and extrapolation to unseen NACA 4-digit geometries, and generalization to unseen non-NACA airfoils. The resulting surrogate accurately captures aerodynamic quantities and qualitative flow features while providing orders-of-magnitude speedups over conventional computational fluid dynamics.
\end{abstract}
\vspace{-10pt}
\section{Introduction}\label{sec:Introduction}
Airfoil aerodynamics~\citep{Anderson1984FundamentalsOA} underpins the performance of a wide range of engineering systems, from aircraft wings~\citep{abbott1959theory} and propellers~\citep{leishman2006principles} to automotive components~\citep{hucho2013aerodynamics,katz2016racecar} designed to generate aerodynamic downforce. In these applications, performance is rarely determined by a single geometry evaluated at a fixed operating condition: engineers must understand how aerodynamic behavior varies across families of airfoil shapes and flow conditions. This makes aerodynamic design an inherently parametric problem.

Despite decades of progress in computational fluid dynamics~(CFD)~\citep{Roache1972CFD,Anderson1995CFD,TuYeohLiu2018CFD}, evaluating airfoil performance across a vast array of possible geometries remains computationally expensive. In particular, CFD can only resolve one discrete configuration at a time. While individual simulations are tractable, systematic exploration across geometry variations quickly becomes prohibitive, particularly in early-stage design where many candidate shapes must be assessed. Surrogate modeling offers a potential path toward alleviating this cost. Data-driven models trained on CFD data can provide rapid predictions once deployed, enabling broader exploration of design spaces. However, many existing surrogates are constructed around fixed geometries or rely on carefully chosen low-dimensional shape parameters. These restrict the range of airfoils that can be represented and make it difficult to generalize beyond the configurations seen during training. Addressing this limitation requires models that can account for geometry variation in a flexible yet physically meaningful way. 

We explore such a representation for steady, two-dimensional subsonic airfoil aerodynamics. We focus on downforce-generating airfoils relevant to road car design, including rear wings, front wings, and active aerodynamic surfaces used to control vehicle stability and performance. We introduce a new dataset with nearly 10,000 samples across 907 different airfoil geometries, and we propose a geometry-aware modeling approach that enables a single surrogate to capture flow behavior across a broad family of airfoil shapes and operating conditions, while remaining computationally efficient. More concretely, we explore different spectral techniques to improve airfoil geometry conditioning, including Fourier-based features from previous computer vision work~\citep{tancik2020fourier}, polymorphic Fourier-based encodings of geospatial objects (Poly2Vec)~\citep{siampoupoly2vec} and our newly proposed Airfoil2Vec decomposition. Note that our geometry conditioning approach is applicable to airfoils in general, but we choose to explore downforce-generating airfoils because these allow us to also open-source a dataset for an underexplored aerodynamic configuration which we believe is interesting for the community.

Our contributions are the following: (I) We introduce a new dataset for steady, incompressible, two-dimensional subsonic flow around downforce-generating NACA 4-digit series airfoils, built from approximately 10,000 Reynolds-Averaged Navier–Stokes (RANS) simulations. (II) We propose geometry-conditioned neural surrogates as fast surrogate models capable of producing flow-field predictions orders of magnitude faster than CFD methods. Our contribution focuses especially on comparing neural fields against neural ODEs (NODEs) and MLPs against graph-based methods over CFD meshes. (III) On top of that, we perform a thorough study comparing different spectral methods for geometry conditioning in the context of airfoil splines and propose a new spectral method for this particular use case called Airfoil2Vec. (IV) We test the generalization capabilities of our model, both in terms of AoA interpolation and geometry interpolation and extrapolation within the NACA 4-digit series, as well as generalization to other non-NACA airfoils used in automotive and motorsport aerodynamics. (V) We evaluate the performance of our final model both in terms of machine learning metrics and through the lens of an aerodynamicist. This provides a comprehensive perspective on model performance, not only in terms of numerical metrics but also in terms of qualitative flow features, which we believe is paramount for interpretability.

\section{Related Work}\label{sec:Related Work}
\paragraph{Reduced Order Modeling.} Reduced-order models (ROMs) lessen the cost of repeated high-fidelity simulations in parametric aerodynamic design by learning compact flow representations that preserve dominant physics while enabling rapid evaluation~\citep{Ripepi2018ReducedorderMF}. They have been applied extensively to unsteady laminar and turbulent flows via coherent-structure extraction and modal decompositions~\citep{Durmaz2013, Stabile2017, Juan2019, Nidhan2020spodwake, Schmidt2018jet, Fukami2021, Giannopoulos2020}. Linear ROMs can perform well even in nonlinear regimes but often depend on real-time or partial state information to offset unmodeled nonlinearities~\citep{Juan2019, Papadakis2021, Loiseau2018, savarino_papadakis_2022}. Nonlinear ROMs are more expressive, yet introduce calibration, robustness, and interpretability challenges~\citep{Nair_Goza_2020, Kim_kim_won_lee_2021, Rozon_Breitsamter_2021}. Many ROMs also remain limited by fixed geometries or low-dimensional shape parameterizations, reducing suitability for geometry-varying airfoil design spaces. Deep learning has become a parallel surrogate-modeling approach in aerospace~\citep{Ling2016ReynoldsAT, Borde2021ConvolutionalNN, multitask, Sayyari2022UnsupervisedDL, Fang2019NeuralNM, Kaandorp2018MachineLF, Li2022MachineLI, Xu2021MachineLF, Li2021OnDG, Liu2023DeeplearningbasedAS, LozanoDuran2020SelfcriticalMW, WaxeneggerWilfing2021MachineLM}. These surrogates can learn highly nonlinear aerodynamic input-output mappings directly from data.

\paragraph{Airfoil Datasets.} The rise of data-driven surrogate models for airfoil (and wing~\citep{Catalani2024NeuralFF}) aerodynamics has been enabled by an expanding set of public CFD datasets that vary in motivation, fidelity, and target users: some were built as ML benchmarks, while others began as engineering databases and were later adopted for ML. ML-focused resources include AirfRANS~\citep{bonnet2022airfrans}, which provides RANS solutions for NACA 4- and 5-digit airfoils across Reynolds-number and angle-of-attack ranges, and UniFoil~\citep{kanchi2025unifoil}, a large-scale collection of roughly half a million RANS simulations spanning laminar, transitional, and turbulent regimes over diverse geometries and operating conditions, intended as a broadly applicable benchmark with explicit transitional coverage. Engineering-first datasets now widely used in ML include the Open Energy Data Initiative airfoil CFD collections~\citep{ramos2023airfoil2k}, offering thousands of geometries evaluated over multiple angles of attack and Reynolds numbers with both flow fields and integrated coefficients, and the Politecnico di Milano RANS flow-field dataset~\citep{schillaci2021airfoilml}, which provides dense solutions for thousands of NACA airfoils at fixed conditions and has been used for field reconstruction and surrogate modeling. High-fidelity databases further expand the landscape, notably PALMO~\citep{cornelius2025palmo}, which uses NASA’s OVERFLOW solver to generate tens of thousands of simulations for NACA 4-series airfoils across wide Reynolds-number and angle-of-attack ranges, including subsonic and transonic regimes, making it valuable for training and validation. 

\section{CFD Data Generation}\label{sec:CFD Data Generation}
\begin{wrapfigure}{r}{0.48\textwidth}
    \centering
    \includegraphics[width=\linewidth]{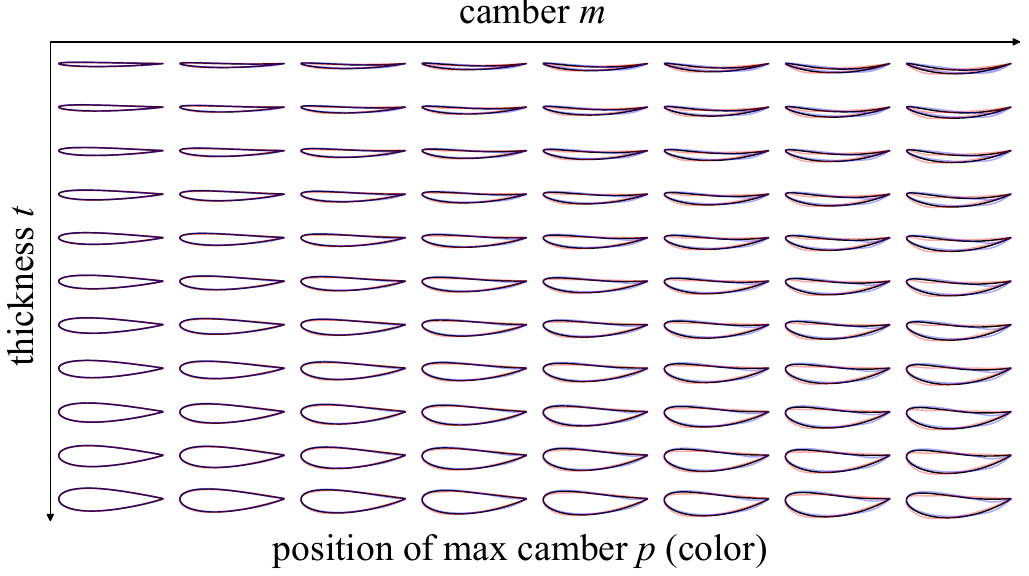}
    \caption{Example of a few inverted (negative camber) NACA 4-digit airfoils parameterized by camber $m$, position of maximum camber $p$, and thickness $t$.}
    \label{fig:airfoils}
\end{wrapfigure}

We aim to develop the first dataset of downforce-generating airfoils. Here, we briefly describe the dataset generation pipeline. Further details on the airfoil geometry, meshing, boundary conditions, and solver configuration are provided in Appendix~\ref{app:Additional CFD Generation Details}. To generate a controlled and interpretable dataset, we adopt the NACA 4-digit family as a parametric baseline. While modern Formula 1 wings involve highly optimized multi-element geometries, inverted single-element airfoils capture the fundamental mechanisms of downforce generation. The NACA 4-digit series provides smooth, analytically defined profiles parameterized by maximum camber $m$, camber position $p$, and thickness $t$, enabling systematic geometric variation. In this study, negative camber ($m<0$) and a sharp trailing-edge are imposed to obtain downforce-oriented sections. We sweep the integer parameters of the designation NACA$MPTT$, $M\in\{1,\dots,8\}$ (maximum camber, \% of chord), $P\in\{1,\dots,8\}$ (position of maximum camber, tenths of chord), and $T\in\{4,\dots,18\}$ (thickness, \% of chord), yielding 960 candidate geometries; the corresponding chord fractions are $m=-M/100$, $p=P/10$ and $t=T/100$ (Appendix~\ref{app:Additional CFD Generation Details}). Of these, 907 produce physically valid airfoils and are retained. Example geometries are shown in Figure~\ref{fig:airfoils}. For each airfoil, steady simulations are performed over 11 angles of attack in the range $\alpha\in[-10^{\circ},0^{\circ}]$, resulting in a total of 9,977 CFD cases.

All simulations are conducted using steady incompressible RANS with the $k-\omega$ SST turbulence model in OpenFOAM. Flow conditions correspond to $Re=4.66\times10^6$ and $M_\infty=0.2$, representative of a clean, straight-line operating condition. A structured C-type mesh is generated automatically for each geometry. A grid-independence study performed on NACA6412 indicates convergence of aerodynamic coefficients at approximately 177k cells; to ensure robustness across the full geometry range, a finer mesh with 276k cells is adopted (Figure~\ref{fig:mesh} in Appendix~\ref{app:Additional CFD Generation Details}). Wall-adjacent resolution falls within the recommended range for wall-modeled RANS simulations. In total, the dataset comprises 9,977 converged simulations across 907 geometries, requiring approximately $1.6\times10^4$ CPU-hours.

\section{Spectral Geometry-Conditioning of Airfoils for Neural Surrogate Models}\label{sec:Spectral Geometry-Conditioning of Airfoils for Neural Surrogate Models}
As previously mentioned, in this work we compare neural fields and neural ODEs (NODEs), as well as using multilayer perceptrons (MLPs) vs graph neural network (GNN) methods for each of them --the quintessential geometric deep learning architecture~\citep{gnn,gdl,borde2025mathematicalfoundationsgeometricdeep}. MLPs are standard feedforward networks which perform predictions pointwise on the mesh coordinates (the nodes of the mesh grid/graph), whereas GNNs learn signals over graphs. For GNNs, message-passing on layer $l$ over a graph $G$ is computed as $
x_{i}^{(l+1)}=\phi\Big(x_{i}^{(l)},\bigoplus_{j\in\mathcal{N}(v_{i})}\psi(x_{i}^{(l)},x_{j}^{(l)})\Big),$ where $\bigoplus$ is a permutation-invariant aggregator over the node $v_i$'s local first-hop neighborhood which is used to compute messages combining node features $x_{i}^{(l)}$ and $x_{j}^{(l)}$ (note that here we use $(l)$ superscript to reference layers, whereas later we use superscripts for referring to points on the airfoil boundary, do not confuse the notation). Both $\psi$ and $\phi$ are non-linear functions. The main motivation behind using GNNs is that they can help make the flow field prediction less sensitive to high frequency noise due to their local averaging nature, but at the same time this can lead to complications capturing sharp flow features~\citep{borde2026learning}.

Each airfoil is parameterized as a spline and discretized into \(N=2,000\) chordwise-aligned points (1,000 per surface) \(\mathcal{A}=\{(x_i^a, y_i^a)\}_{i=1}^N\). The points are sampled using a sinusoidal distribution in the chordwise coordinate \(x\) with spacing \(\Delta x\in[10^{-6},1.6\times10^{-3}]c\), producing an ordered sequence that captures the camber and thickness distributions and the leading- and trailing-edge profiles, with additional point clustering near the leading and trailing edges. A geometry encoder \(\mathcal{E}_g\) maps this sequence to a fixed-dimensional latent representation $g = \mathcal{E}_g(\mathcal{A}) \in \mathbb{R}^{d_g}$. Conditioned on spatial coordinates \((x,y)\), angle of attack (AoA) \(\alpha\), and the geometry embedding \(g\), the neural field predicts the local (pointwise) flow quantities, pressure and velocity components, via $(p, v_x, v_y) = f_1(x, y, \alpha, g).$

\paragraph{Fourier-based features.} Artificial neural networks operating directly on raw spatial coordinates often exhibit a spectral bias toward low-frequencies and can struggle to represent sharp spatial gradients~\citep{tancik2020fourier}, such as those occurring in boundary layers and wake regions~\citep{borde2023aerothermodynamicsimulatorsrocketdesign}. To mitigate this limitation, one might apply a Fourier feature mapping to the spatial coordinates \(\mathbf{x} = (x, y)\): $\Gamma(\mathbf{x}) =
    \begin{bmatrix}
    \cos(2\pi \mathbf{B}\mathbf{x}) \\
    \sin(2\pi \mathbf{B}\mathbf{x})
    \end{bmatrix}
    \in \mathbb{R}^{2l},
    \quad \mathbf{B} \sim \mathcal{N}(0, \sigma^2).$ Here, \(\mathbf{B} \in \mathbb{R}^{l \times 2}\) is a random projection matrix whose entries are drawn independently from a Gaussian distribution with variance \(\sigma^2\). The number of frequencies \(l\) and bandwidth parameter \(\sigma\) are selected empirically. Our primary baseline (in the neural field case acting pointwise on the flow field coordinates for simplicity here) therefore takes the form $(p, v_x, v_y) = f_2(\Gamma(x,y), \alpha, g),$
where \(f_2\) is an MLP trained jointly with the geometry encoder \(\mathcal{E}_g\) in an end-to-end manner. Fourier-based features also apply to GNNs and NODEs (they only modify the network input). Note that this is a \textit{local positional encoding}: different points in the flow domain get different Fourier features: $(x_i,y_i) \rightarrow \Gamma(x_i,y_i)$. In simple terms, its purpose is essentially: \textit{At what spatial location should I predict the flow?}

In the NODE models, the conditioned input is first mapped to a latent state $h(0)=E_{\mathrm{in}}(\Gamma(x,y),\alpha,g)$ and then evolved over an artificial depth variable $t\in[0,1]$ according to $\dot h=f_\theta(h,t;\alpha,g)$. The final state $h(1)$ is mapped to $(p,v_x,v_y)$ through shared prediction heads. The ODE is solved numerically (Runge--Kutta) with a fixed number of integration steps.

\paragraph{Contour-level Poly2Vec.} The Fourier features introduced above encode spatial coordinates pointwise, but do not provide a spectral representation of the airfoil boundary as a geometric object. The goal of Poly2Vec is instead to encode \textit{which airfoil is generating the flow}. It therefore provides a \textit{global geometry representation of the airfoil} and does not compete with the pointwise Fourier features introduced above; rather, the two representations are complementary.

To make this distinction precise, we interpret the ordered sequence $\mathcal{A}=\{(x_i^a,y_i^a)\}_{i=1}^{N}$ as a closed, piecewise-linear airfoil boundary obtained by joining consecutive points. Writing $\mathbf a_i=(x_i^a,y_i^a)^{\mathsf T}$, we close the contour by defining $\mathbf a_{N+1}=\mathbf a_1$. Before computing its spectral representation, the contour can optionally be placed in a canonical chord-based coordinate system. Let $c_a=\max_i x_i^a-\min_i x_i^a$ denote the airfoil chord length and, assuming that the boundary is ordered from trailing edge to leading edge and back to the trailing edge, let $y_{\mathrm{TE}}^a=(y_1^a+y_N^a)/2$ denote the vertical location of the trailing-edge midpoint. The normalized coordinates are $\widetilde{x}_i^a
    =
    \frac{x_i^a-\min_j x_j^a}{c_a},
    \qquad
    \widetilde{y}_i^a
    =
    \frac{y_i^a-y_{\mathrm{TE}}^a}{c_a}.$
This removes differences due to global translation and uniform scaling while deliberately preserving the chordwise orientation of the airfoil. In many aerodynamic datasets, airfoil coordinates are already expressed in nondimensional chord coordinates, in which case this normalization is largely redundant; we nevertheless state it explicitly as an optional preprocessing step to ensure a common origin and scale across datasets. The transformation does not remove rotation, which is intentional: the airfoil is kept in a canonical chordwise orientation while the angle of attack $\alpha$ is supplied separately to the surrogate. For notational simplicity, we subsequently drop the tildes.

Let $\boldsymbol{\omega}=(u,v)^{\mathsf T}\in\mathbb{R}^2$ denote the spatial-frequency vector, where $u$ and $v$ are Fourier-conjugate to the chordwise coordinate $x^a$ and vertical coordinate $y^a$, respectively. If the airfoil coordinates are nondimensionalized by the chord length, $u$ and $v$ may be interpreted as frequencies in cycles per chord along the corresponding directions. For an individual boundary point $(x_i^a,y_i^a)$, represented as a Dirac measure, the Fourier transform is $F_i(u,v)=\exp[-2\pi\mathrm{i}(ux_i^a+vy_i^a)]$, which has unit magnitude at every frequency. Thus, a pointwise transform encodes the location of each sampled boundary point through phase, but its magnitude contains no information about geometric extent. Poly2Vec instead constructs a spectral representation of the complete contour by analytically aggregating the contributions of the line segments joining consecutive boundary points. For segment $j=1,\ldots,N$, let its endpoints be $\mathbf q_j=\mathbf a_j$ and $\mathbf r_j=\mathbf a_{j+1}$, its displacement $\mathbf d_j=\mathbf r_j-\mathbf q_j$, its midpoint $\mathbf m_j=(\mathbf q_j+\mathbf r_j)/2$, and its length $\ell_j=\|\mathbf d_j\|_2$. Let $L_{\mathcal A}=\sum_{j=1}^{N}\ell_j$ denote the total contour length. The normalized contour spectrum is then
\begin{equation}
    Z_{\mathcal A}(\boldsymbol{\omega})
    =
    \frac{1}{L_{\mathcal A}}
    \sum_{j=1}^{N}
    \ell_j
    \exp\!\left[
        -2\pi\mathrm{i}\,
        \boldsymbol{\omega}^{\mathsf T}\mathbf m_j
    \right]
    \operatorname{sinc}\!\left(
        \boldsymbol{\omega}^{\mathsf T}\mathbf d_j
    \right),
    \label{eq:contour-cft}
\end{equation}
where $\operatorname{sinc}(z)=\sin(\pi z)/(\pi z)$ with $\operatorname{sinc}(0)=1$. Writing $\mathbf m_j=(m_{x,j},m_{y,j})^{\mathsf T}$ and $\mathbf d_j=(d_{x,j},d_{y,j})^{\mathsf T}$, the two inner products are $\boldsymbol{\omega}^{\mathsf T}\mathbf m_j=u m_{x,j}+v m_{y,j}$ and $\boldsymbol{\omega}^{\mathsf T}\mathbf d_j=u d_{x,j}+v d_{y,j}$. Equation~\ref{eq:contour-cft} is therefore the Fourier transform of the normalized arc-length measure supported on the piecewise-linear airfoil boundary. 

The original Poly2Vec line-segment construction uses a squared-length prefactor $\ell_j^2$. For airfoil contours we instead use the arc-length weight $\ell_j$, which makes the representation invariant to arbitrary subdivision of a straight segment. For example, splitting a segment of length $\ell$ into two equal parts preserves the linear weight, $\ell=\ell/2+\ell/2$, whereas squared-length weighting changes from $\ell^2$ to $2(\ell/2)^2=\ell^2/2$. This is desirable because the sampling density of an airfoil boundary is a numerical choice rather than an intrinsic property of the geometry. Normalization by $L_{\mathcal A}$ removes the trivial multiplicative dependence of the spectrum on total contour length: it gives $Z_{\mathcal A}(\mathbf 0)=1$ and, by the triangle inequality, $|Z_{\mathcal A}(\boldsymbol{\omega})|\leq1$ for all $\boldsymbol{\omega}\in\mathbb{R}^2$. Without this normalization, the zero-frequency coefficient is exactly $L_{\mathcal A}$. In practice, the continuous spectrum $Z_{\mathcal A}(u,v)$ is evaluated on a finite set of spatial frequencies, whose range and resolution are treated as encoder hyperparameters; full details are provided in Appendix~\ref{app:frequency-sampling}.

\begin{figure}[hbpt!]
    \centering
    \includegraphics[width=0.95\linewidth,trim=0 0 0 20,
    clip]{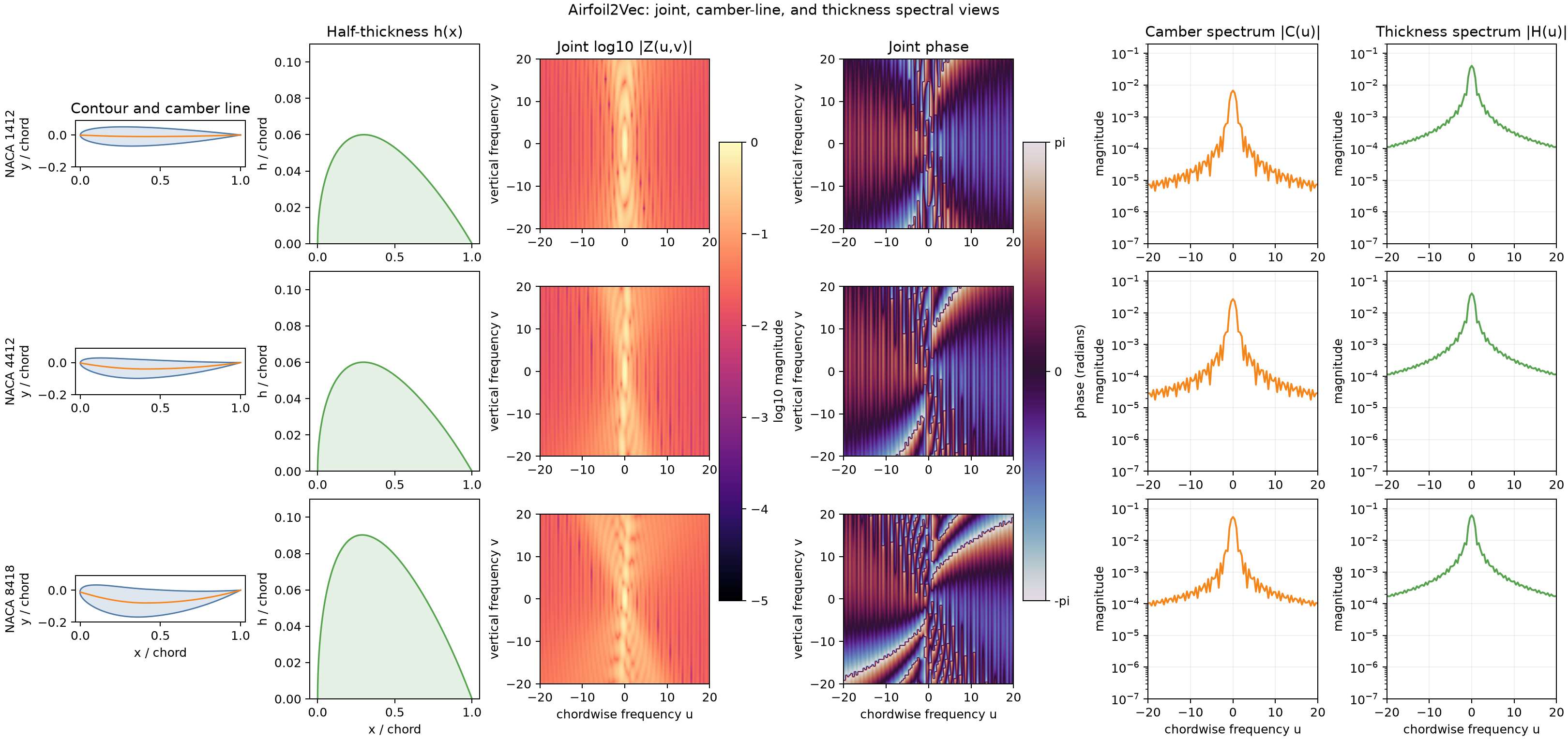}
    \caption{Complementary spectral views used by Airfoil2Vec for representative airfoils.}
    \label{fig:airfoil2vec-3view}
\end{figure}

\paragraph{Airfoil2Vec.}
Contour-level Poly2Vec provides a generic spectral representation of the complete airfoil boundary through $Z_{\mathcal A}(\boldsymbol{\omega})=Z_{\mathcal A}(u,v)$, but it treats $\mathcal A$ as a generic closed contour. Airfoils possess additional structure: after chordwise alignment, their upper and lower surfaces are functions of a common normalized chordwise coordinate $x^a\in[0,1]$. We propose \textit{Airfoil2Vec} encodings, which exploit this structure while \emph{retaining Poly2Vec as one of its spectral views}. In particular,
\begin{equation}
    \operatorname{Airfoil2Vec}(\mathcal A)
    =
    \left\{
    Z_{\mathcal A}(u,v),\,
    C_{\mathcal A}(u),\,
    H_{\mathcal A}(u)
    \right\},
    \label{eq:airfoil2vec}
\end{equation}
where $Z_{\mathcal A}(u,v)$ is the contour-level Poly2Vec spectrum defined in Equation~\ref{eq:contour-cft}, while $C_{\mathcal A}(u)$ and $H_{\mathcal A}(u)$ explicitly encode the airfoil camber and thickness distributions, respectively. Let $y_u^a(x^a)$ and $y_l^a(x^a)$ denote the upper and lower surfaces of $\mathcal A$ evaluated at the same chordwise coordinate $x^a$. We define the camber line as $c_{\mathcal A}(x^a)=[y_u^a(x^a)+y_l^a(x^a)]/2$ and the half-thickness as $h_{\mathcal A}(x^a)=[y_u^a(x^a)-y_l^a(x^a)]/2$. The original surfaces can therefore be reconstructed exactly as $y_u^a(x^a)=c_{\mathcal A}(x^a)+h_{\mathcal A}(x^a)$ and $y_l^a(x^a)=c_{\mathcal A}(x^a)-h_{\mathcal A}(x^a)$. This decomposition also clarifies how camber and thickness enter the joint Poly2Vec spectrum. Ignoring, for intuition only, both the line-segment integration factors and the unequal arc-length weights of the upper and lower surfaces, the paired Fourier contribution of the upper and lower surfaces at a fixed $x^a$ is
\begin{align}
z_{\mathcal A}(x^a;u,v)
&=
\exp\!\left[-2\pi\mathrm{i}
\left(u x^a+v y_u^a(x^a)\right)\right]
+
\exp\!\left[-2\pi\mathrm{i}
\left(u x^a+v y_l^a(x^a)\right)\right]
\nonumber\\
&=
2
\exp\!\left[-2\pi\mathrm{i}
\left(u x^a+v c_{\mathcal A}(x^a)\right)\right]
\cos\!\left(2\pi v h_{\mathcal A}(x^a)\right).
\label{eq:paired-contribution}
\end{align}
Equation~\ref{eq:paired-contribution} is an illustrative paired-point identity rather than an exact decomposition of $Z_{\mathcal A}$: in the full spectrum (Equation~\ref{eq:contour-cft}) each segment carries its own arc-length weight and integration factor, which differ between the two surfaces, so the separation below holds only approximately. It nevertheless motivates the airfoil-specific decomposition. Camber enters the joint spectrum through the complex phase, while thickness modulates its response through the cosine term. The chordwise frequency $u$ additionally captures where these variations occur along the chord. Thus, the generic Poly2Vec spectrum $Z_{\mathcal A}$ contains both camber and thickness information, but mixes them within a single two-dimensional representation. Airfoil2Vec retains this complete coupled representation while exposing the two physically meaningful factors separately. The additional Airfoil2Vec views are the one-dimensional Fourier transforms of the camber and half-thickness distributions,
$C_{\mathcal A}(u)=\int_0^1 c_{\mathcal A}(x^a)\exp(-2\pi\mathrm{i}u x^a)\,\mathrm{d}x^a$
and
$H_{\mathcal A}(u)=\int_0^1 h_{\mathcal A}(x^a)\exp(-2\pi\mathrm{i}u x^a)\,\mathrm{d}x^a$.
The camber spectrum $C_{\mathcal A}$ makes the magnitude and chordwise variation of camber directly available to the geometry encoder, while $H_{\mathcal A}$ analogously represents the thickness distribution. Although $c_{\mathcal A}$ and $h_{\mathcal A}$ are sufficient to reconstruct the two airfoil surfaces, retaining $Z_{\mathcal A}$ remains useful because it directly preserves their nonlinear coupling in the full two-dimensional contour geometry rather than requiring the downstream network to recover these interactions from the separated representations. Figure~\ref{fig:airfoil2vec-3view} visualizes these complementary views. The joint magnitude and phase correspond to the Poly2Vec component $Z_{\mathcal A}(u,v)$, while the camber and thickness spectra correspond to the two additional Airfoil2Vec views $C_{\mathcal A}(u)$ and $H_{\mathcal A}(u)$.

\begin{figure}[hbpt!]
    \centering
    \includegraphics[width=0.95\linewidth,trim=0 0 0 20,clip]{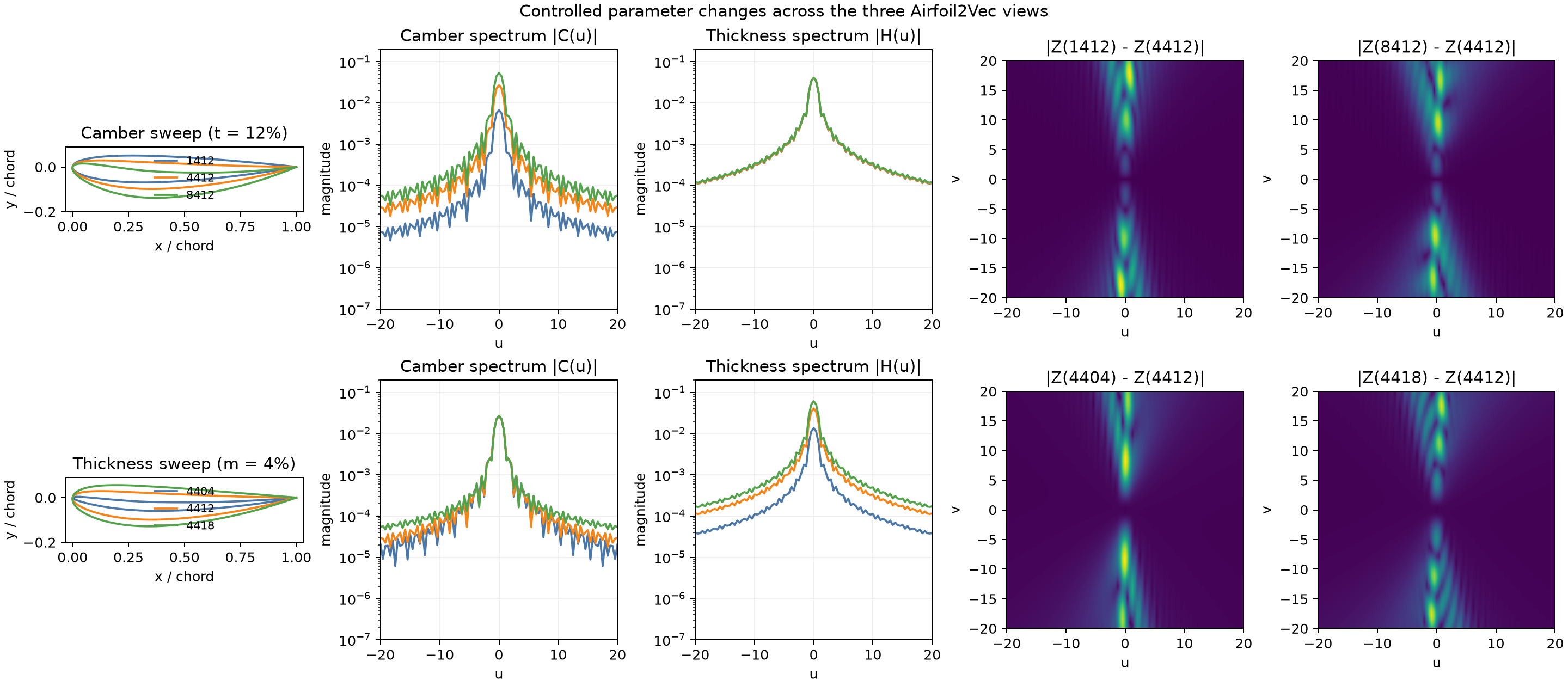}
    \caption{Controlled camber and thickness variations across the Airfoil2Vec spectral views. In the camber sweep, variations in the airfoil predominantly modify $C_{\mathcal A}$ while the thickness representation remains comparatively stable; in the thickness sweep, the converse behavior is observed for $H_{\mathcal A}$. The joint $Z_{\mathcal A}$ responds to both types of variation, motivating the additional separated Airfoil2Vec views.}
    \label{fig:airfoil2vec-sweep}
\end{figure}

In Figure~\ref{fig:airfoil2vec-sweep}, varying camber while keeping thickness approximately fixed predominantly changes $C_{\mathcal A}$, whereas varying thickness at fixed camber predominantly changes $H_{\mathcal A}$. The joint spectrum $Z_{\mathcal A}$ from Poly2Vec responds to both types of variation. The chordwise extent of an airfoil is substantially larger than its vertical extent, reflecting the strongly anisotropic nature of the geometry and, correspondingly, of its spectral content.

Additional details about the geometry encoder architecture are provided in Appendix~\ref{app:geometry-encoders}. Full details of the frequency grid and its hyperparameters are provided in Appendix~\ref{app:frequency-sampling}.

\section{Quantitative Results} 
\label{sec:Quantitative Results}

\paragraph{Training procedure.} All models are trained under identical conditions. Every geometry is simulated at the same 11 AoAs, $\alpha\in\{0^\circ,-1^\circ,\dots,-10^\circ\}$, and the $9{,}977$ flow fields are partitioned into two disjoint sets. The \emph{training set} contains $8{,}600$ flow fields: 860 geometries, each at 10 AoAs (all except $\alpha=-5^\circ$). The \emph{validation set} contains the remaining $1{,}377$ flow fields, organized into three subsets: (i)~the \emph{AoA interpolation} subset, i.e., the $\alpha=-5^\circ$ flow fields of the 860 training geometries (860 flow fields); (ii)~the \emph{geometry interpolation} subset, 10 geometries of moderate camber and thickness withheld at all AoAs (110 flow fields); and (iii)~the \emph{geometry extrapolation} subset, 37 geometries of high camber combined with very thin or thick sections, withheld at all AoAs (407 flow fields). Hence, the training and validation sets share no flow field, but only subsets (ii) and (iii), 47 geometries in total, are geometry-disjoint from the training set (Appendix~\ref{app:training procedure}, Table~\ref{tab:data_split}). The validation set is used for early stopping, checkpoint selection and all model comparisons in this section; ``validation MSE'' denotes the MSE averaged over all $1{,}377$ validation flow fields. No separate test split is reserved: throughout the paper, a \emph{test} denotes an evaluation study on one of the three validation subsets (e.g., the AoA interpolation test of Section~\ref{sec:Aerodynamic analysis}). Inputs and outputs are min--max normalized to $[0,1]$ and AoA $\alpha$ is scaled by $1/10$. During training we randomly subsample $4{,}096$ cells per flow field to form fixed-size batches. The model is trained to minimize mean squared error over $(p,v_x,v_y)$ using AdamW (learning rate $10^{-3}$, weight decay $10^{-2}$) with cosine annealing to $10^{-6}$ over 100 epochs, gradient clipping at 1.0, and mixed-precision BF16 on 8 H100 GPUs. We employ early stopping with patience 10 on the validation loss. For more details, see Appendix~\ref{app:training procedure}.

\subsection{Fourier mappings for flow-field prediction}

\paragraph{Effect of spatial Fourier features.} We first evaluate whether Fourier mappings are useful for learning the pressure and velocity fields. As shown in Table~\ref{tab:fourier_baseline}, adding Fourier features to the spatial coordinates gives a substantial improvement over using raw coordinates alone. The validation MSE decreases from $1.08{\times}10^{-3}$ to $1.77{\times}10^{-4}$, corresponding to roughly a $6\times$ improvement. This suggests that the Fourier mapping helps the network represent higher-frequency flow features that are poorly captured by a standard coordinate MLP, in line with previous work~\citep{borde2023aerothermodynamicsimulatorsrocketdesign,borde2026learning}. Additional hyperparameter sweeps for Fourier bandwith and dimension can be found in Appendix~\ref{appendix:bandwith_dim}.

\begin{table}[hbpt!]
    \centering
    \caption{Validation MSE (mean $\pm$ std over 3 seeds) for baseline vs.\ Fourier encoding.}
    \label{tab:fourier_baseline}
    \begingroup
    \setlength{\tabcolsep}{4pt}
    \small
    \scalebox{0.7}{
    \begin{tabular}{@{}l c c c c@{}}
        \toprule
        Encoding & Val.\ MSE & MAE $p$ & MAE $v_x$ & MAE $v_y$ \\
        \midrule
        No Fourier & $1.08{\times}10^{-3}$ $\pm$ $3.92{\times}10^{-4}$ & 0.0222 $\pm$ 0.0049 & 0.0221 $\pm$ 0.0034 & 0.0147 $\pm$ 0.0017 \\
        Fourier ($\sigma{=}45, l=256$) & $\mathbf{1.77{\times}10^{-4}}$ $\pm$ $\mathbf{1.29{\times}10^{-5}}$ & 0.0138 $\pm$ 0.0008 & 0.0119 $\pm$ 0.0005 & 0.0106 $\pm$ 0.0010 \\
        \bottomrule
    \end{tabular}}
    \endgroup
\end{table}

\begin{wraptable}{r}{0.35\textwidth}
    \centering
    \caption{Validation MSE for different graph neighbourhood sizes $k$ at $\sigma=45$ ($\times 10^{-4}$).}
    \label{tab:k_results}
    \scalebox{0.7}{
    \begin{tabular}{lccccc}
        \toprule
        $k$ & 2 & 4 & 6 & 8 & 10 \\
        \midrule
        GCN     & \textbf{2.17} & 2.37 & 2.59 & 2.54 & 2.75 \\
        GAT     & \textbf{2.10} & 2.26 & 2.41 & 2.55 & 2.70 \\
        GATv2   & \textbf{2.41} & 2.63 & 2.78 & 2.91 & 3.08 \\
        SGC     & \textbf{2.02} & 2.18 & 2.31 & 2.47 & 2.71 \\
        ChebNet & \textbf{1.98} & 2.07 & 2.16 & 2.30 & 2.55 \\
        GIN     & \textbf{2.13} & 2.24 & 2.38 & 2.56 & 2.82 \\
        \bottomrule
    \end{tabular}}
\end{wraptable}

\subsection{Comparing neural fields, graph models, and neural ODEs}
 
\paragraph{Graph Neural Networks.} We construct an undirected $k$NN graph over mesh cell centers and evaluate six GNN operators: GCN \citep{gcn}, GAT \citep{gat}, GATv2 \citep{gatv2}, SGC \citep{sgc}, ChebNet \citep{chebnet}, and GIN \citep{gin}. The models are trained with the same inputs, prediction heads, and compute budget as the neural field. Sweeping $k\in\{2,4,6,8,10\}$ (the number of node neighbours) at $\sigma=45$ shows that $k=2$ performs best for all six operators, with performance generally deteriorating as the neighborhood grows (see Table~\ref{tab:k_results}). We therefore use $k=2$ for the direct GNN baselines; for the graph-based NODEs below, $k$ is swept jointly with the number of integration steps (Table~\ref{tab:ode_ablation_gnn}).

Table~\ref{tab:gnn_fourier} shows that Fourier encoding is also important for graph-based models (when we drop them there is a clear drop in performance). Across all six GNN models, adding the spatial Fourier mapping reduces validation MSE by roughly $3.5$--$4\times$. Among the GNNs, ChebNet gives the best validation MSE and is competitive with the neural field, although it does not improve over it in this direct comparison. Based on these results, there does not appear to be a benefit to using GNNs over MLPs for neural fields, which is consistent with previous work~\citep{borde2026learning}.

\begin{table}[hbpt!]
    \centering
    \caption{Val loss (mean $\pm$ std over 3 seeds) for GNN architectures ($k=2$) with and without Fourier encoding.}
    \label{tab:gnn_fourier}
    \begingroup
    \setlength{\tabcolsep}{4pt}
    \small
    \scalebox{0.6}{
    \begin{tabular}{@{}l c c c c@{}}
        \toprule
        Model 
        & Val.\ MSE ($\times 10^{-4}$) 
        & MAE $p$ ($\times 10^{-2}$) 
        & MAE $v_x$ ($\times 10^{-2}$) 
        & MAE $v_y$ ($\times 10^{-2}$) \\
        \midrule
        GCN (No Fourier) 
        & $8.17 \pm 0.31$ 
        & $1.57 \pm 0.16$ 
        & $2.33 \pm 0.34$ 
        & $1.48 \pm 0.23$ \\

        GCN (Fourier) 
        & $\mathbf{2.16 \pm 0.06}$ 
        & $\mathbf{1.39 \pm 0.04}$ 
        & $\mathbf{1.24 \pm 0.12}$ 
        & $\mathbf{1.35 \pm 0.12}$ \\
        \midrule

        GAT (No Fourier) 
        & $7.99 \pm 0.28$ 
        & $1.78 \pm 0.12$ 
        & $2.01 \pm 0.21$ 
        & $1.55 \pm 0.18$ \\
        
        GAT (Fourier) 
        & $\mathbf{2.10 \pm 0.07}$ 
        & $\mathbf{1.33 \pm 0.05}$ 
        & $\mathbf{1.73 \pm 0.14}$ 
        & $\mathbf{0.90 \pm 0.06}$ \\
        \midrule

        GATv2 (No Fourier) 
        & $8.87 \pm 0.55$ 
        & $1.82 \pm 0.50$ 
        & $2.12 \pm 0.66$ 
        & $1.69 \pm 0.64$ \\

        GATv2 (Fourier) 
        & $\mathbf{2.46 \pm 0.21}$ 
        & $\mathbf{1.56 \pm 0.18}$ 
        & $\mathbf{1.55 \pm 0.26}$ 
        & $\mathbf{1.00 \pm 0.15}$ \\

        \midrule

        SGC (No Fourier) 
        & $7.72 \pm 0.34$ 
        & $1.54 \pm 0.14$ 
        & $2.18 \pm 0.22$ 
        & $1.47 \pm 0.18$ \\

        SGC (Fourier) 
        & $\mathbf{2.03 \pm 0.08}$ 
        & $\mathbf{1.31 \pm 0.05}$ 
        & $\mathbf{1.22 \pm 0.10}$ 
        & $\mathbf{1.11 \pm 0.09}$ \\

        \midrule

        ChebNet (No Fourier) 
        & $7.41 \pm 0.27$ 
        & $1.49 \pm 0.11$ 
        & $2.07 \pm 0.19$ 
        & $1.41 \pm 0.17$ \\

        ChebNet (Fourier) 
        & $\mathbf{1.95 \pm 0.07}$ 
        & $\mathbf{1.23 \pm 0.04}$ 
        & $\mathbf{1.15 \pm 0.09}$ 
        & $\mathbf{0.97 \pm 0.08}$ \\

        \midrule

        GIN (No Fourier) 
        & $8.03 \pm 0.42$ 
        & $1.67 \pm 0.21$ 
        & $2.21 \pm 0.27$ 
        & $1.55 \pm 0.24$ \\

        GIN (Fourier) 
        & $\mathbf{2.11 \pm 0.09}$ 
        & $\mathbf{1.34 \pm 0.06}$ 
        & $\mathbf{1.30 \pm 0.12}$ 
        & $\mathbf{1.08 \pm 0.10}$ \\
        \bottomrule
        
    \end{tabular}}
    \endgroup
\end{table}

\begin{wraptable}{r}{0.35\textwidth}
    \centering
    \caption{MLP-NODE val MSE for different numbers of integration steps (mean $\pm$ std over 3 seeds).}
    \label{tab:integration_steps}
    \scalebox{0.7}{
    \begin{tabular}{cc}
        \toprule
        \# Integration Steps & Validation MSE $\times 10^{-4}$ \\
        \midrule
        2  & $1.74 \pm 0.24$ \\
        4  & $1.53 \pm 0.09$ \\
        8  & $\mathbf{1.52 \pm 0.22}$ \\
        16 & $1.80 \pm 0.59$ \\
        32 & $1.84 \pm 0.41$ \\
        \bottomrule
    \end{tabular}}
\end{wraptable}

\paragraph{Neural ODEs.} Table~\ref{tab:integration_steps} reports the integration-step sweep for the MLP-NODE. Performance is best at $8$ steps, reaching a validation MSE of $1.52\times10^{-4}$. Using fewer or more steps does not improve accuracy. The MLP-NODE improves on the neural field and all GNN baselines reported above. We also test graph-based NODEs by sweeping both graph connectivity and integration steps for each GNN operator (Table~\ref{tab:ode_ablation_gnn}). Because GNN-NODEs perform message passing at every integration step, their cost grows quickly with both graph connectivity and integration depth. We therefore restrict this sweep to relatively sparse graphs and low-step configurations. The preferred hyperparameters vary across architectures, although sparse graphs and few integration steps perform well in most cases.

\begin{table*}[hbpt!]
    \centering
    \caption{
    GNN-NODE hyperparameter ablation.
    For the graph-connectivity sweep (left), the number of integration steps is fixed at $s=4$.
    For the integration-step sweep (right), graph connectivity is fixed to the best value
    $k^\star$ found for each model in the connectivity sweep.
    Validation MSE is reported in units of $\times 10^{-4}$.
    }
    \label{tab:ode_ablation_gnn}
    \scalebox{0.6}{
    \begin{tabular}{lcccc c c cccc}
        \toprule
        & \multicolumn{4}{c}{Connectivity sweep ($s=4$ fixed)}
        & &
        \multicolumn{5}{c}{Integration-step sweep ($k=k^\star$ fixed)} \\
        \cmidrule(lr){2-5}
        \cmidrule(lr){7-11}
        Model
        & $k=1$ & $k=2$ & $k=4$ & $k=8$
        &&
        $k^\star$
        & $s=1$ & $s=2$ & $s=4$ & $s=6$ \\
        \midrule

        GCN
        & 1.60 & 1.58 & 1.51 & \textbf{1.26}
        &&
        8
        & 1.35 & 1.44 & 1.23 & \textbf{1.21} \\

        GAT
        & \textbf{1.09} & 1.17 & 1.23 & 1.34
        &&
        1
        & \textbf{1.05} & 1.34 & 1.16 & 1.45 \\

        GATv2
        & \textbf{1.20} & 1.28 & 1.30 & 1.58
        &&
        1
        & \textbf{1.25} & 1.26 & 1.28 & 1.32 \\

        SGC
        & \textbf{1.25} & 1.37 & 1.28 & 1.34
        &&
        1
        & \textbf{1.22} & 1.49 & 1.41 & 1.52 \\

        ChebNet
        & 1.30 & \textbf{1.25} & 1.45 & 1.47
        &&
        2
        & \textbf{1.29} & 1.42 & 1.49 & 1.55 \\

        GIN
        & 1.38 & 1.28 & \textbf{1.15} & 1.44
        &&
        4
        & 1.21 & \textbf{0.82} & 1.60 & 1.62 \\

        \bottomrule
    \end{tabular}}
\end{table*}

\begin{table*}[htbp!]
    \centering
    \caption{Final validation MSE for the selected GNN-NODE configuration of each GNN operator. Each column reports the selected graph connectivity $k$ and number of integration steps in parentheses. Results are mean $\pm$ standard deviation over 3 seeds ($\times 10^{-4}$).}
    \label{tab:final_results}
    \scalebox{0.6}{
    \begin{tabular}{lcccccc}
        \toprule
        GNN 
        & GAT 
        & GATv2 
        & GCN 
        & GIN 
        & SGC 
        & ChebNet \\
        & $(k{=}1, s{=}1)$ 
        & $(k{=}1, s{=}1)$ 
        & $(k{=}8, s{=}6)$ 
        & $(k{=}4, s{=}2)$ 
        & $(k{=}1, s{=}1)$ 
        & $(k{=}2, s{=}1)$ \\
        \midrule
        Validation MSE
        & $1.19 \pm 0.19$
        & $1.19 \pm 0.26$
        & $1.35 \pm 0.06$
        & $1.09 \pm 0.26$
        & $1.42 \pm 0.18$
        & $\mathbf{1.07 \pm 0.11}$ \\
        \bottomrule
    \end{tabular}}
\end{table*}

\subsection{Ablation of geometry encoding methods}\label{subsec:geometry-encoding-ablation}

We ablate only the geometry encoder, keeping the spatial Fourier mapping $\Gamma(x,y)$ for the flow-field coordinates fixed. Airfoil2Vec achieves the best performance for both the neural field and GNN-NODE, with the latter also giving the strongest result overall. This supports the benefit of explicitly encoding airfoil geometry through its camber--thickness structure. The exception is MLP-NODE, for which Poly2Vec performs best (note that Poly2Vec is technically contained within Airfoil2Vec, so the model may not be leveraging the additional Fourier transforms, and this could only be true for this particular dataset size).

\begin{wraptable}{r}{0.35\textwidth}
\centering
\caption{Geometry-encoding ablation. Validation MSE
($\times 10^{-4}$), mean $\pm$ std over 3 seeds. GNN-NODE uses ChebNet
($k=2$, steps $=1$).}
\label{tab:ablation-all}
\scalebox{0.58}{
\begin{tabular}{lccc}
\toprule
Configuration & MLP-NF & GNN-NODE & MLP-NODE \\
\midrule
Baseline & $1.77 \pm 0.13$ & $1.07 \pm 0.11$ & $1.52 \pm 0.22$ \\
Poly2Vec & $1.48 \pm 0.49$ & $1.03 \pm 0.07$ & $\mathbf{1.42 \pm 0.17}$ \\
Airfoil2Vec & $\mathbf{1.41 \pm 0.14}$ & $\mathbf{0.98 \pm 0.05}$ & $1.94 \pm 0.14$ \\
\bottomrule
\end{tabular}}
\end{wraptable}

\subsection{Neural simulation time} 

Neural fields are the cheapest to query, while NODE variants incur numerical integration costs. We measure full-field inference time for pressure and velocity components over all 276k mesh cells for one airfoil and angle of attack on an NVIDIA H100 80GB GPU at a time. Models use \texttt{torch.compile} with PyTorch 2.9 default settings; timings exclude data loading and one-off compilation. Averaged over 40 airfoils, the neural field takes 280--290 ms for all three geometry encoders (baseline, Poly2Vec, and Airfoil2Vec), with differences within GPU noise. The MLP-NODE takes 2.26 s and the GNN-NODE about 5.49 s for all encoders. Standard deviations are below 10 ms in every case. Geometry encoding therefore has negligible impact on inference time, as the geometry spectra are computed once per airfoil and runtime is dominated by the flow network. The GNN-NODE is slowest because graph message passing is repeated at each integration step. By comparison, each OpenFOAM CFD simulation takes about 48 minutes (Figure~\ref{fig:simplefoam_performance}, and could be further accelerated with more CPU cores), making even the slowest surrogate roughly $500\times$ faster. Hand-optimized CUDA kernels are left for future work.

\section{Aerodynamic analysis}\label{sec:Aerodynamic analysis}
We evaluate the predictions of the Airfoil2Vec-conditioned models from an aerodynamics perspective, focusing first on flow fields around NACA 4-digit series airfoils. The normalized percentage error is defined as $\mathcal{E}_i = 100\cdot\,\frac{\lvert q_{i,t} - q_{i,p}\rvert}{\max\!\big(\lvert q_{i,t}\rvert + \lvert q_{i,p}\rvert,\ \tau\big)}\hspace{4pt}[\%],$ where $q_{i,t}$ is the true state, $q_{i,p}$ the approximated state and $\tau$ a small positive threshold keeping the error bounded when the states are near zero. To quantify the overall error on the actual data points (not only on the interpolated fields), we also consider the $\mathcal{L}_2$-norm based error $\mathcal{L}_{2,i} = 100\cdot\frac{\parallel q_{i,t} - q_{i,p} \parallel_2}{\parallel q_{i,t} \parallel_2 + \parallel q_{i,p} \parallel_2} \hspace{4pt}[\%],$ reported for each state variable in Table~\ref{tab:l2_tests}. Three tests showcase the predictive accuracy of the models, each analyzing one case drawn from the corresponding subset of the validation set (Section~\ref{sec:Quantitative Results}): AoA interpolation (NACA4207, trained on every AoA of the polar except the withheld $\alpha=-5^\circ$), geometry interpolation (NACA4412, withheld at all AoAs, while similar airfoil splines are in the training set) and geometry extrapolation (NACA8604, withheld at all AoAs; camber, position of maximum camber and thickness of the airfoil lie outside the training parameter space). Across the MLP-NF, MLP-NODE and GNN-NODE models, Table~\ref{tab:l2_tests} shows that MLP-NODE is more accurate than MLP-NF for every test and state variable, and so is GNN-NODE with a single exception: $v_x$ in the geometry extrapolation test ($1.09\%$ vs.\ $0.99\%$ for MLP-NF). Overall, the error is almost always $<1\%$ on $v_x$, below 10\% on $v_y$ (except for MLP-NF) and close to zero on $p$. All models thus achieve good predictive accuracy, with MLP-NODE slightly more accurate than GNN-NODE based on the minimum error counts in Table~\ref{tab:l2_tests} (green results).

\begin{table}[hbpt!]
  \centering
  \caption{$\mathcal{L}_2$-norm based error (\%) of $v_x/U_\infty$, $v_y/U_\infty$, $p/p_\infty$ states over the full computational domain for models MLP-NF, MLP-NODE, GNN-NODE. Three tests are performed, one per subset of the validation set: AoA interpolation, geometry interpolation and geometry extrapolation. For each test and state variable, the lowest error across the three models is colored in green.}
  \label{tab:l2_tests}
  \resizebox{0.7\textwidth}{!}{%
  \begin{tabular}{l l c c c c c c c c c c}
    \toprule
    Test & Airfoil & $\alpha$ $[^\circ]$ & \multicolumn{3}{c}{MLP-NF} & \multicolumn{3}{c}{MLP-NODE} & \multicolumn{3}{c}{GNN-NODE} \\
    \cmidrule(lr){4-6} \cmidrule(lr){7-9} \cmidrule(lr){10-12}
    & & & $v_x$ & $v_y$ & $p$ & $v_x$ & $v_y$ & $p$ & $v_x$ & $v_y$ & $p$ \\
    \midrule
    AoA interpolation & NACA4207 & -5 & 0.73 & 7.36 & 0.03 & \textcolor{green}{0.54} & 3.44 & \textcolor{green}{0.02} & 0.59 & \textcolor{green}{3.41} & \textcolor{green}{0.02} \\
    \midrule
    Geometry interpolation & NACA4412 & 0 & 0.66 & 11.68 & 0.03 & \textcolor{green}{0.53} & 8.34 & \textcolor{green}{0.01} & 0.58 & \textcolor{green}{7.93} & 0.02 \\
    \midrule
    Geometry extrapolation & NACA8604 & -5 & 0.99 & 7.31 & 0.04 & \textcolor{green}{0.91} & \textcolor{green}{5.36} & \textcolor{green}{0.03} & 1.09 & 5.81 & \textcolor{green}{0.03} \\
    \bottomrule
  \end{tabular}
  }
\end{table}

The reconstructed fields are shown in Figure~\ref{fig:geom_extrap} for the geometry extrapolation test and in Appendix~\ref{app:AoA-geometry Interpolation Tests} (Figures~\ref{fig:aoa_interp}-\ref{fig:geom_interp}) for the other two. We focus here on geometry extrapolation, the most challenging scenario for the models. The NACA8604 lies outside the parametric range of the training data (NACA$86TT$ airfoils were only included in the training set for $5\le T\le16$, see Table~\ref{tab:data_split}). The MLP-NODE prediction is compared against the CFD truth at $-5^\circ$ incidence. The reconstructed velocity fields show remarkable agreement: most of the error is concentrated in the thin wake for $v_x$, in the trailing-edge separation region on the lower surface for $v_y$, and in the thin boundary layer. The predicted fields nevertheless remain physically consistent and quantitatively comparable to the ground truth, providing engineering-relevant predictions even in this extrapolation regime. The pressure field, whose error is negligible, is shown in Appendix~\ref{app:AoA-geometry Interpolation Tests}.

\begin{figure}
    \centering
    \includegraphics[width=0.8\linewidth]{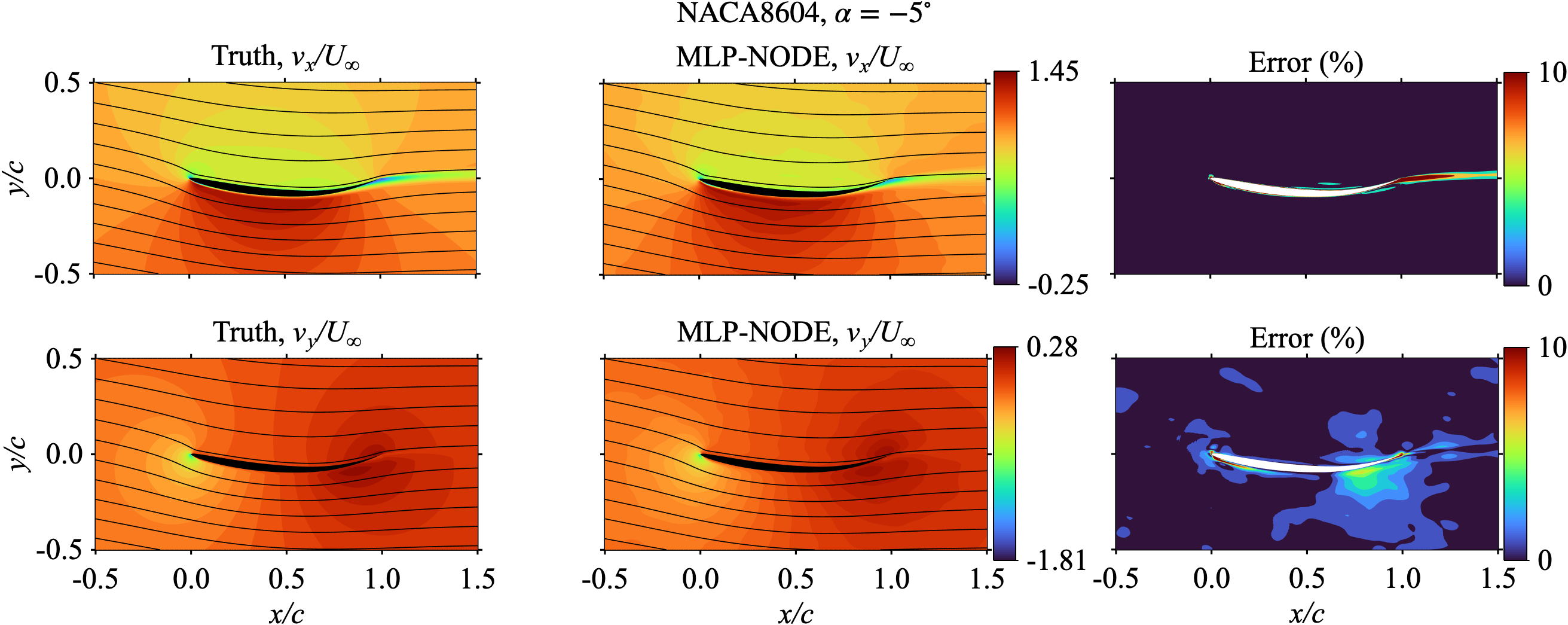}
    \caption{Comparison of (first row) $v_x/U_\infty$ and (second row) $v_y/U_\infty$ between OpenFOAM and the MLP-NODE prediction. This case illustrates the model’s ability to extrapolate airfoil geometry, as the NACA8604 was not included in the training set at any AoA. Overall $\mathcal{L}_2$-norm based error $[\mathcal{L}_{2,v_x},\mathcal{L}_{2,v_y},\mathcal{L}_{2,p}] = [0.91, 5.36, 0.03]\%$.}
    \label{fig:geom_extrap}
\end{figure}

\begin{figure}
    \centering
    \includegraphics[width=\linewidth]{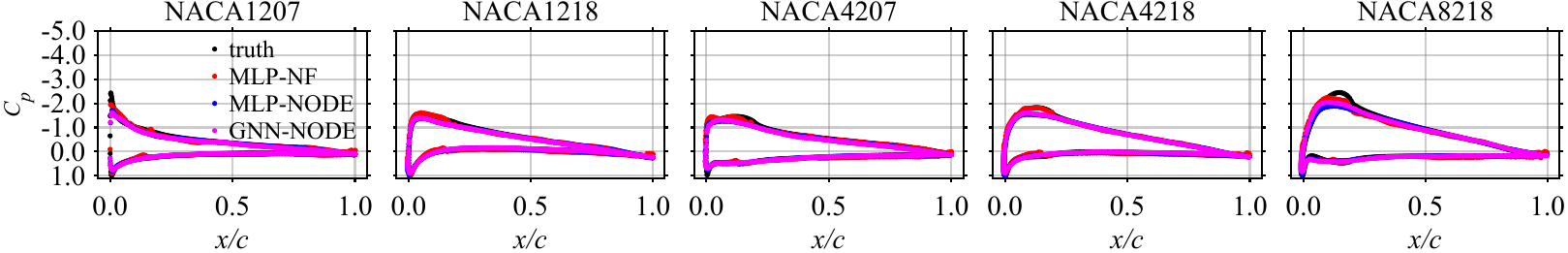}
    \caption{Pressure coefficient at $\alpha=-5^{\circ}$ for a few airfoil geometries demonstrating AoA interpolation capability of the models. The models were trained on each of these geometries at every AoA except the one shown, which belongs to the AoA interpolation subset of the validation set.}
    \label{fig:Cp_AoA_interp}
\end{figure}

Although not visible in the error contours, the predicted velocity states in Figure~\ref{fig:geom_extrap} show residual background noise scattered across the field of view, absent in the CFD solution. Spurious oscillations are more visible in the reconstructed surface pressure coefficient $C_p=\frac{p-p_{\infty}}{0.5\rho U_{\infty}^2}$ in Figure~\ref{fig:Cp_AoA_interp}, extracted for 5 airfoil geometries of variable camber and thickness at $\alpha=-5^\circ$, representative of the models' AoA interpolation. The estimated distributions follow the CFD data: the suction peak (minimum pressure) and the pressure recovery on the lower surface are captured with acceptable discrepancies, while the lower-surface stagnation point ($C_p\approx1$) and overall pressure evolution almost overlap across all models. The MLP-NF is locally oscillatory around the main $C_p$ trend but captures the suction peak better than the NODE models, which dampen sharp gradients while reducing spurious oscillations. Additional $C_p$ comparisons for different airfoils in Appendix~\ref{app: Pressure coefficient interpolation-extrapolation} demonstrate the models' geometry interpolation and extrapolation accuracy.

\begin{figure}
    \centering
    \includegraphics[width=\linewidth]{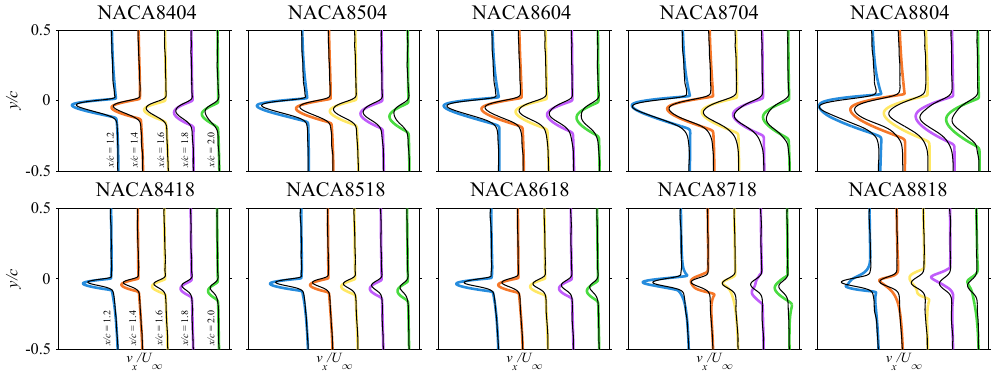}
    \caption{$v_x/U_\infty$ profiles at 5 streamwise stations in the $\alpha=-10^\circ$ wake of NACA airfoils with 8\% camber and 4\%-18\% thickness representative of thin and thick cambered airfoils. Truth: solid colored lines. MLP-NODE prediction: solid black lines.}
    \label{fig:wake_airfoil}
\end{figure}

\paragraph{Analyzing characteristic flow features.} Two observations deriving from figures~\ref{fig:geom_extrap}-\ref{fig:Cp_AoA_interp} are the error concentrated in the wake region and the noise in the reconstructed flow field. We investigate these aspects next. We select a group of airfoils belonging to the geometry extrapolation subset of the validation set and extract streamwise velocity profiles across the wake to assess the wake (deficit) reconstruction. Two sets of sample airfoils are selected for the investigation: an 8\%-cambered, 4\%-thick set and an 8\%-cambered, 18\%-thick set. The position of maximum camber (the second digit $P$ of the designation NACA$MPTT$) is varied from 4/10 of the chord to 8/10 of the chord. The wake profiles are therefore shown in Figure~\ref{fig:wake_airfoil} at $\alpha=-10^\circ$. The MLP-NODE model predictions plotted in this figure are in good agreement with the CFD data. Across the board of airfoils shown, the model generally under-predicts the streamwise velocity deficit, which could be ascribed to the ``diffusivity'' of NODE models. This is especially accentuated in aft-cambered airfoils, i.e. NACA8804, NACA8818, etc. In the thick airfoil set, the model also misses the exact shape of the wake, especially at locations farther downstream in the wake ($x/c\geq1.8$). The wake thickness and velocity deficit are captured more faithfully by the model in the thin airfoil set. Supplementary results reporting the $\mathcal{L}_2$-norm error of the reconstructed streamwise velocity field $v_x/U_\infty$ in a patch residing in the wake of a group of airfoils are provided in Appendix~\ref{app:wake prediction} and tabulated for each model and the full angle of attack polar, including basic statistics to quickly assess the best model for this task.

\paragraph{Spectral noise analysis of the model prediction.} Figure~\ref{fig:spectral_noise} shows the three spectral diagnostics of Appendix~\ref{sec:Spectral Noise analysis} for $v_y/U_\infty$ of NACA8604 at $\alpha=-5^\circ$, in a farfield patch and a near-wake patch (settings and derived scales in Table~\ref{tab:psd_settings}; high-frequency residuals for all states in Table~\ref{tab:noise_metric}). In the farfield patch the truth is empty above the cutoff, so the reading is unambiguous: every model places energy at scales the flow does not have ($E_p/E_t\gg1$, $\gamma^2\approx0$), which cannot stem from attenuating, amplifying or displacing the truth (Appendix~\ref{sec:Spectral Noise analysis}), i.e. pure noise, and its level ranks the models. MLP-NF is the noisiest, with about $2.6\times$ the noise RMS of MLP-NODE; the two NODE models are comparable, MLP-NODE slightly lower for $v_y$ and GNN-NODE for $v_x$ and $p$.
\begin{figure}[t]
    \centering
    \includegraphics[width=\linewidth]{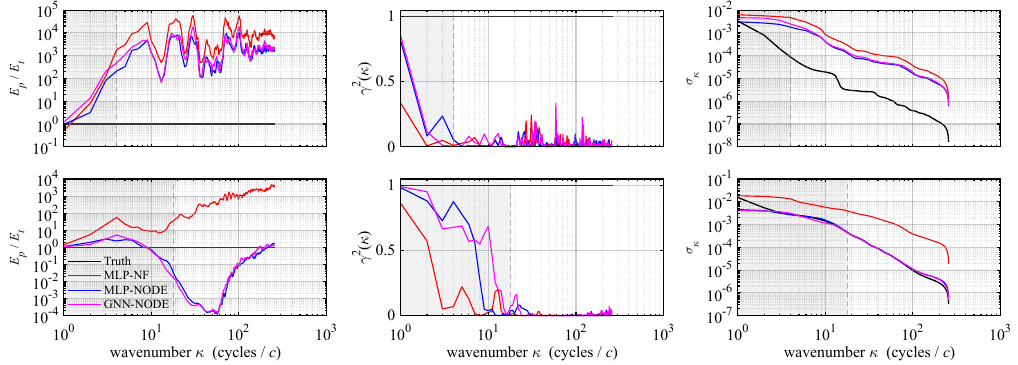}
    \caption{Spectral noise analysis of $v_y/U_\infty$ for NACA8604 at $\alpha=-5^\circ$ (Appendix~\ref{sec:Spectral Noise analysis}): prediction-to-truth energy ratio $E_p/E_t$, spectral coherence $\gamma^2(\kappa)$ and cumulative tail RMS $\sigma_\kappa$ versus wavenumber $\kappa$ (cycles/$c$) in the (top) farfield patch and (bottom) near-wake patch. The gray band marks $\kappa<\kappa_c$, excluded from the high-frequency residual $\sigma_{\text{hf}}$.}
    \label{fig:spectral_noise}
\end{figure}
In the near-wake patch the truth is populated down to the mesh scale and the models separate in kind rather than in degree. The NODE models remain coherent with the truth at large scales; below $\lambda\approx0.1c$ their prediction-to-truth ratio falls far under unity and $\sigma_\kappa$ collapses onto the truth tail, so they smooth the shear layer scales out rather than adding noise ($\sigma_{\text{hf}}$ at or below the omission bound and $\sigma_{\text{inc}}\ll\sigma_{\text{hf}}$ in Table~\ref{tab:noise_metric}), GNN-NODE holding coherence to about twice the wavenumber of MLP-NODE. MLP-NF instead loses coherence already at wake-width scales ($\lambda\approx0.3$-$0.5c$) while its ratio rises above unity there and keeps growing towards the Nyquist limit: the small-scale wake content is replaced by spurious energy with noise on top, and its $\sigma_{\text{hf}}$ exceeds that of the NODE models by an order of magnitude. Whether the loss of coherence at wake-width scales reflects a mispredicted wake thickness, velocity deficit or orientation cannot be decided from the spectra alone, which measure energy per scale but not where it sits, and must be checked against wake profiles and contours. The $v_x$ profiles at $\alpha=-10^\circ$ for the NACA8604 airfoil (Fig.~\ref{fig:wake_airfoil}) are well reproduced by MLP-NODE, which suggests that for the NODE models the large-scale discrepancy is a mild amplitude excess rather than a topological error; for MLP-NF the low coherence hints at a genuinely different cross-stream wake structure, a point that the $v_y$ profiles would have to confirm.

\paragraph{Generalization to non-NACA airfoils.} To assess generalization beyond the training distribution, we evaluate the three models on five non-NACA airfoils commonly used or studied in automotive and motorsport aerodynamics (these are not part of the dataset: they belong to neither the training nor the validation set and were not used for early stopping or model selection): NASA LS(1)-0413, Selig S1223, Eppler E423, Clark Y, and Wortmann FX 63-137. These profiles span diverse aerodynamic design characteristics relevant to downforce generation and probe extrapolation to geometries with substantially different camber distributions, thickness profiles, and aerodynamic characteristics compared to the NACA 4-digit splines seen during training. Despite lying outside the NACA training distribution, the models generalize well, achieving low streamwise velocity $\mathcal{L}_2$-norm error across the full AoA range, with mean error below $3\%$ for all five geometries (Table~\ref{tab:l2_non_naca}), and the predicted flow fields also closely reproduce the corresponding CFD solutions, as illustrated in Figure~\ref{fig:non_naca_mlp} in Appendix~\ref{app:non-NACA airfoils}.

\section{Conclusion}\label{sec:Conclusion}
Our results show that geometry-conditioned neural surrogates can accurately reproduce steady aerodynamic flow fields while reducing prediction time by $500\times$ compared to our CFD simulations. Fourier spatial features consistently improve performance. NODEs outperform standard neural fields and plain GNNs but at the cost of inference time. Airfoil2Vec is empirically shown to be a strong conditioning technique for airfoil geometries. Importantly, the models generalize well to unseen NACA geometries and non-NACA airfoils, with pressure fields reconstructed particularly accurately, despite the CFD dataset being relatively small compared to large-scale AI datasets found in natural language processing and computer vision. The largest errors arise in wake velocity prediction, especially for more challenging extrapolation regimes, rearward camber, and larger incidence. Our results highlight spectral geometry conditioning as an effective tool for fast aerodynamic surrogate modeling.

\newpage
\bibliography{tmlr}
\bibliographystyle{tmlr}

\clearpage
\appendix

\section{Additional CFD Generation Details}\label{app:Additional CFD Generation Details}
In this appendix we provide an extended discussion regarding the CFD data generation pipeline.

While modern Formula 1 wings employ highly customized multi-element profiles optimized through CFD and wind-tunnel testing, downforce-generating devices in motorsport are fundamentally inverted wings~\citep{Katz2006AERODYNAMICSOR}.
In automotive aerodynamic studies, NACA 4-digit sections are frequently used as convenient baselines (e.g., NACA4412~\citep{Maddani2024SelectingTO} and NACA6412~\citep{Chiplunkar2022ComputationalFD}). Motivated by this context, we adopt the NACA 4-digit series as a simplified and well-defined baseline family. These profiles provide smooth, analytically defined geometries that are easy to parameterize and modify, making them suitable for automated numerical studies. The goal is not to reproduce contemporary F1 designs, but to explore downforce-oriented airfoil behavior within a controlled and interpretable shape family.

\paragraph{NACA 4-digit series geometry generation.} The NACA 4-digit series is fully defined by 3 parameters, encoded as integers in the designation NACA$MPTT$: the maximum camber $M$ in percent of the airfoil's chord (i.e. the distance from the leading-edge to the trailing-edge), the position of maximum camber $P$ in tenths of the chord, and the thickness $T$ in percent of the chord. For instance, the NACA4412 airfoil has $M=4$ (maximum camber equal to 4\% of the chord), $P=4$ (maximum camber located at a streamwise coordinate equal to 4/10 of the chord) and $T=12$ (12\% thickness). The analytical expressions below require the corresponding chord fractions, which for the inverted (downforce-generating) geometries of this work are
\begin{equation}
    m=-\frac{M}{100},\qquad p=\frac{P}{10},\qquad t=\frac{T}{100},
    \label{eq:naca_parameters}
\end{equation}
so that, e.g., the inverted NACA4412 has $m=-0.04$, $p=0.4$ and $t=0.12$. With $x\in[0,1]$ the chordwise coordinate normalized by the chord, the mean camber line of cambered NACA 4-digit airfoils is defined analytically as

\begin{equation}
    y_c(x)=
    \begin{cases}
    \displaystyle \frac{m}{p^2}\left(2px-x^2\right), & 0\le x \le p,\\[8pt]
    \displaystyle \frac{m}{(1-p)^2}\left((1-2p)+2px-x^2\right), & p < x \le 1.
    \end{cases}
    \label{eq:camber_line}
\end{equation}

The thickness distribution is given by
\begin{equation}
    y_t(x)=5t\left(
    0.2969\sqrt{x}-0.1260x-0.3516x^2+0.2843x^3-0.1036x^4
    \right).
    \label{eq:thickness}
\end{equation}
Provided these two functions, the upper and lower surface coordinates can be defined as
\begin{subequations}\label{eq:airfoil_coordinates}
    \begin{align}
    x_U(x) &= x - y_t(x)\sin\theta(x), \label{eq:airfoil_coordinates_xU}\\
    y_U(x) &= y_c(x) + y_t(x)\cos\theta(x), \label{eq:airfoil_coordinates_yU}\\[6pt]
    x_L(x) &= x + y_t(x)\sin\theta(x), \label{eq:airfoil_coordinates_xL}\\
    y_L(x) &= y_c(x) - y_t(x)\cos\theta(x). \label{eq:airfoil_coordinates_yL}
    \end{align}
\end{subequations}
where $\theta(x)$ is computed from the derivative (the slope) of the mean camber line $\theta(x)=\arctan\!\left(\frac{dy_c}{dx}(x)\right)$. Note that in our application we opt for a sharp trailing-edge (achieved by imposing the coefficient of the 4th-order term in equation \ref{eq:thickness} to -0.1036, as opposed to -0.1015 for a square-type trailing-edge) and negative camber, $m=-M/100<0$ in Equation~\ref{eq:naca_parameters}, to obtain downforce-type airfoils. A few representative geometries from the generated dataset are shown in Figure~\ref{fig:airfoils}. Thin and thick, low-, medium- and high-camber, fore- and aft-cambered airfoils are obtained. For the present dataset, 960 NACA 4-digit codes are considered by setting $M\in\{1,\dots,8\}$, $P\in\{1,\dots,8\}$, and $T\in\{4,\dots,18\}$ ($8\times8\times15$), of which 907 result in physically meaningful airfoil geometries; the remaining cases lead to malformed shapes that cause CFD divergence, and we discard them.

\begin{figure}[hbt!]
\centering
\begin{subfigure}{0.32\linewidth}
  \centering
  \includegraphics[width=0.95\linewidth]{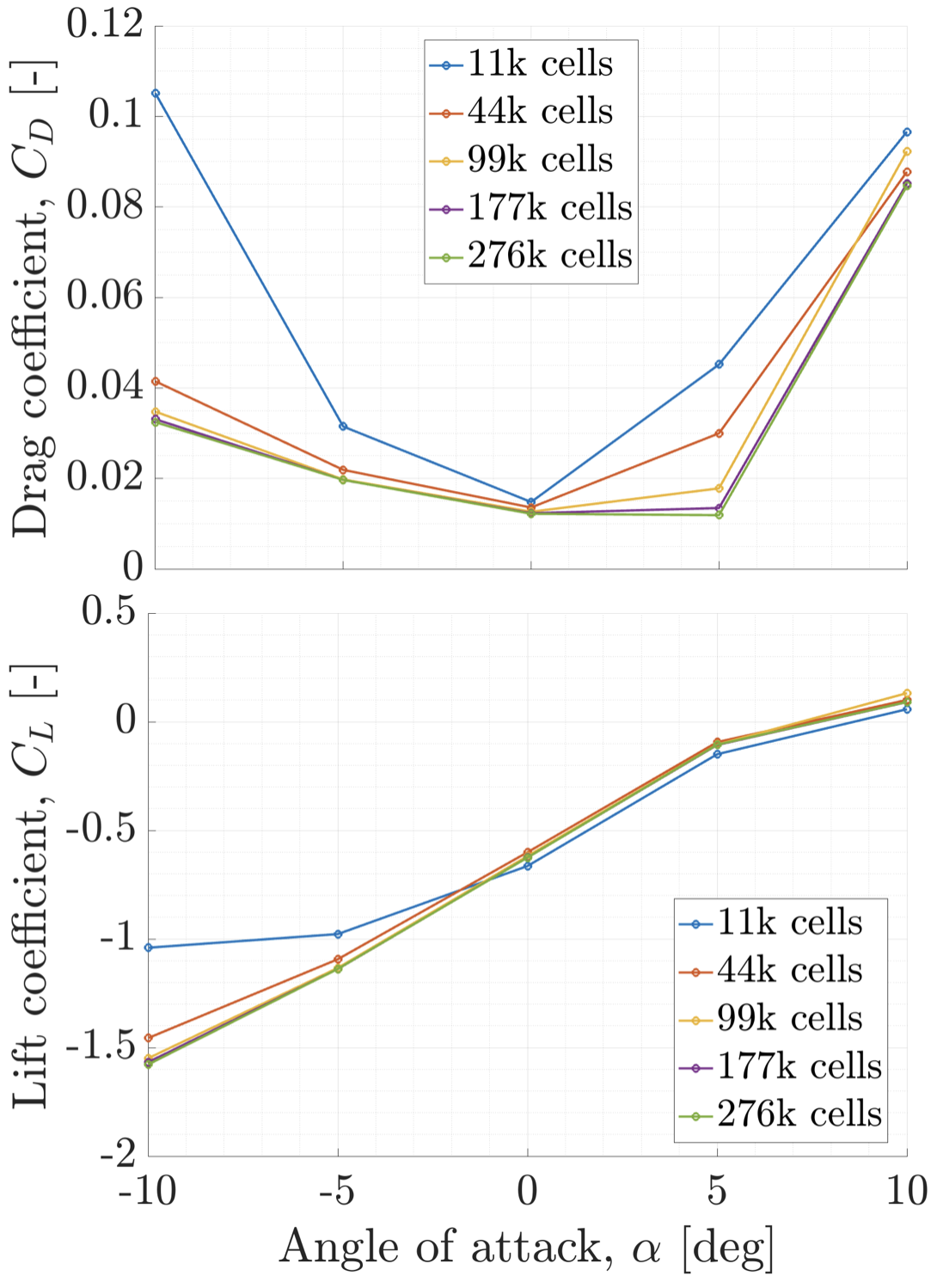}
  \caption{Mesh study.}
  \label{fig:mesh_study}
\end{subfigure}
\hspace{0.01cm}
\begin{subfigure}{0.66\linewidth}
  \centering
  \includegraphics[width=1\linewidth]{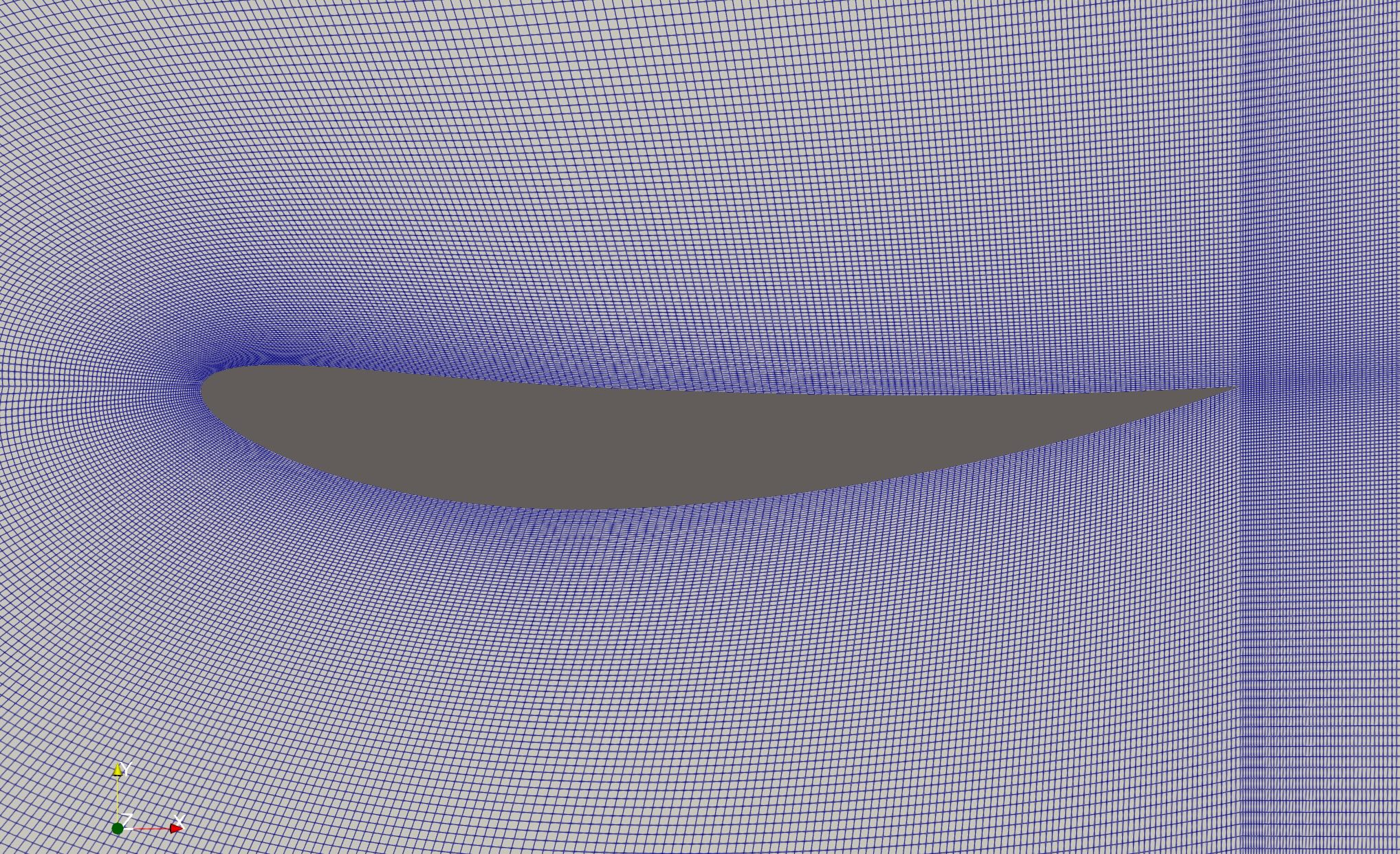}
  \caption{Selected mesh (276k cells).}
  \label{fig:mesh_final}
\end{subfigure}%
\caption{Grid-independence study and selected mesh for the NACA6412 airfoil at $M_{\infty} = 0.2$ and $Re = 4.66 \times 10^6$.}
\label{fig:mesh}
\end{figure}

\paragraph{Domain size and mesh generation.} A C-type 2-D structured computational grid is generated by running a custom-made Python script that writes a $\texttt{blockMeshDict}$ dictionary to execute in OpenFOAM (v2406). The script asks the user to select the NACA 4 digits in the format $MPTT$, the chord length $c$ ($=1$), the number of points to discretize each airfoil surface $n$ ($=1,000$, i.e., 2,000 points in total), the camber sign (positive/negative, negative in our case), the trailing-edge type (square/sharp, sharp in our case), and the domain size in terms of upstream length expressed as the number of chords upstream of the leading-edge located at $(x,y)=(0,0)$ ($=11c$), downstream length taken from the trailing-edge ($=20c$), distance of the top and bottom boundaries taken from $y=0$ ($=12c$). Although the simulation is 2-D, OpenFOAM requests a spanwise length input ($=0.3c$) that effectively sets the out-of-plane cell thickness. A grid-independence study is performed for the NACA6412 airfoil at $M_{\infty} = 0.2$ and $Re = 4.66 \times 10^6$, in which progressively finer meshes are simulated at several angles of attack and the lift and drag coefficients are compared. \autoref{fig:mesh_study} shows that the aerodynamic coefficients converge for the mesh with 177k cells. However, since the simulations in this study span a wide range of airfoil geometries, a finer mesh with 276k cells is adopted to provide a more conservative resolution. The selected mesh is shown in \autoref{fig:mesh_final}. The wall-adjacent cells are characterized by $30 < y^+ < 200$ (on average $y^{+}\approx120$) over most of the surface (with a few exceptions at the stagnation point), which lies within the recommended range for wall-modeled simulations to predict wall shear stresses. 907 meshes are automatically generated by running $\texttt{blockMesh}$ serially on a 128 CPU machine, distributing one mesh run per CPU. The mesh generation with the present settings is very fast, in the order of minutes for ~900 geometries.

\paragraph{Solver and boundary conditions.} Steady-state incompressible RANS simulations are performed with the \texttt{simpleFoam} solver using the $k-\omega$ SST turbulence model. The chosen operational flow conditions are idealistic assumptions for an F1 car front wing under clean/undisturbed conditions in a straight track segment and are tabulated in Table~\ref{tab:flow_conditions}. The velocity $(v_x,v_y)$ states are initialized with $(U_{\infty}\cos \alpha,U_{\infty}\sin \alpha)$, where $U_{\infty}$ is the free-stream velocity and $\alpha$ is the AoA, which is varied in the range $[-10^{\circ},0^{\circ}]$. The turbulent kinetic energy $k=3/2(U_{\infty}I)^2$ and specific dissipation rate $\omega=k^{0.5}/(0.09^{0.25}l)$ are calculated from an assumed freestream turbulence intensity of $I=1\%$ and a turbulent reference length of $l=1\%c$. These conditions may well differ from real F1 scenarios but are physical assumptions for low/medium free-stream turbulence cases. At the inlet and outlet boundaries of the domain we impose free-stream conditions on all quantities, and no-slip velocity and zero-gradient for pressure at the airfoil wall. Wall functions \texttt{kqRWallFunction} and \texttt{omegaWallFunction} are used for $k$ and $\omega$ at the wall.

\begin{table}[hbpt!]
    \centering
    \caption{Flow conditions.}
    \label{tab:flow_conditions}
    \resizebox{0.65\linewidth}{!}{%
    \begin{tabular}{l l}
        \toprule
        Ratio of specific heats, $\gamma$ & 1.4 \\
        Specific gas constant, $R$ ($\mathrm{J\,kg^{-1}\,K^{-1}}$) & 287.1 \\
        Free-stream velocity magnitude, $U_{\infty}$ ($\mathrm{m\,s^{-1}}$) & 68.06 \\
        Free-stream density, $\rho_{\infty}$ ($\mathrm{kg\,m^{-3}}$) & 1.225 \\
        Free-stream temperature, $T_{\infty}$ ($\mathrm{K}$) & 288.15 \\
        Free-stream pressure, $p_{\infty}$ ($\mathrm{Pa}$) & 101341.63 \\
        Free-stream dynamic viscosity, $\mu_{\infty}$ ($\mathrm{kg\,m^{-1}\,s^{-1}}$) & $1.789\times10^{-5}$ \\
        Free-stream kinematic viscosity, $\nu_{\infty}$ ($\mathrm{m^{2}\,s^{-1}}$) & $1.461\times10^{-5}$ \\
        Free-stream speed of sound, $a_{\infty}$ ($\mathrm{m\,s^{-1}}$) & 340.32 \\
        Reynolds number, $Re$ & $4.66\times10^{6}$ \\
        Mach number, $M_{\infty}$ & 0.2 \\
        \bottomrule
    \end{tabular}%
    }
\end{table}

\begin{figure}[hbpt!]
    \centering
    \includegraphics[width=\linewidth]{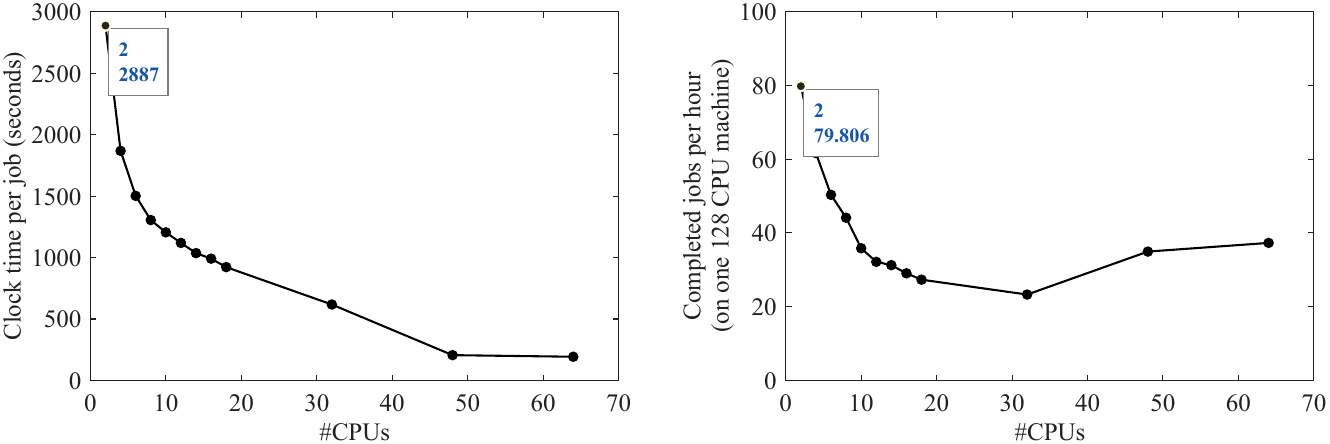}
    \caption{Performance of \texttt{simpleFoam} on the 276,000 cells mesh. (Left) Clock time per job in seconds vs. number of CPUs for parallel execution. (Right) Completed jobs per hour on a 128 CPU machine. The optimal set-up to complete the largest number of experiments in one hour is highlighted with the boxes.}
    \label{fig:simplefoam_performance}
\end{figure}

\paragraph{Performance.} In total, 9,977 CFD run cases (or experiments) are executed, consisting of 907 airfoil geometries and 11 AoAs for each geometry. The \texttt{simpleFoam} solver is run in parallel mode using MPI and 10,000 iterations to converge the residuals to $10^{-7}$ for the velocities, $k$ and $\omega$ and to $10^{-3}$ for pressure, on average. The simulations are distributed across three 128-CPU machines, and \texttt{simpleFoam} is run using two CPUs per case, as this configuration maximizes the number of completed simulations per hour (see \autoref{fig:simplefoam_performance}). $79\times3$ jobs are completed each hour, which means about 42 hours in total to complete 9,977 CFD experiments. Overall, approximately $1.6\times10^4$ CPU-hours ($9{,}977$ cases $\times$ $\approx48$ minutes $\times$ 2 CPUs) were required for this dataset. The RAM usage is very modest in this application, with only 1.5GB per experiment and occupied memory storage 1.2GB per geometry, totaling 1,088GB for the entire OpenFOAM output.

\section{Additional Methodological Details}\label{sec:methodological details}
In this section we provide additional details regarding the architecture used for conditioning. Each of the spectra $Z_{\mathcal A}$, $C_{\mathcal A}$, and $H_{\mathcal A}$ is complex-valued. Rather than supplying the wrapped phase $\arg F$ directly to the learned encoder, for any sampled complex spectrum $F$ we use the real-valued representation
$\Phi(F)=[\log(\varepsilon+|F|),\,\operatorname{Im}F,\,\operatorname{Re}F]$,
where $\varepsilon>0$ is a small numerical constant. Since $\operatorname{Re}F=|F|\cos(\arg F)$ and $\operatorname{Im}F=|F|\sin(\arg F)$, these channels preserve periodic phase information without introducing the artificial $-\pi/\pi$ discontinuity associated with a directly represented wrapped phase. Moreover, the real and imaginary components naturally approach zero when the spectral magnitude is small. The three deterministic spectral views are finally mapped through lightweight learned projections, $\mathbf g_Z=P_Z(\Phi(Z_{\mathcal A}))$, $\mathbf g_C=P_C(\Phi(C_{\mathcal A}))$, and $\mathbf g_H=P_H(\Phi(H_{\mathcal A}))$, and fused into a fixed-dimensional geometry embedding $\mathbf g
    =
    P_{\mathrm{fuse}}
    \left(
    [m_Z\mathbf g_Z,\,
     m_C\mathbf g_C,\,
     m_H\mathbf g_H]
    \right),
    \qquad
    m_Z,m_C,m_H\in\{0,1\}.$ Here, $P_Z$, $P_C$, and $P_H$ denote learned projection networks for the joint, camber, and thickness spectra, respectively, while $P_{\mathrm{fuse}}$ maps their concatenation to the final geometry embedding $\mathbf g\in\mathbb R^{d_g}$. The binary masks $m_Z$, $m_C$, and $m_H$ are used only for controlled ablations: a removed spectral component is replaced by zeros rather than changing the projection or fusion architecture. Contour-level Poly2Vec therefore corresponds to $(m_Z,m_C,m_H)=(1,0,0)$, whereas the complete Airfoil2Vec representation uses $(1,1,1)$. This makes the Poly2Vec--Airfoil2Vec comparison directly test whether explicitly exposing camber and thickness provides useful information beyond the joint contour spectrum while keeping the learned spectral encoder architecture fixed.

\subsection{Poly2Vec Frequency Sampling}\label{app:frequency-sampling}
Although the contour Fourier transform $Z_{\mathcal A}(u,v)$ is defined continuously over $(u,v)\in\mathbb{R}^2$, it must be evaluated at a finite set of frequencies in practice. We use geometrically spaced positive frequencies,
\begin{equation}
    u_k
    =
    u_{\min}
    \left(
    \frac{u_{\max}}{u_{\min}}
    \right)^{\frac{k}{n_u-1}},
    \qquad
    k=0,\ldots,n_u-1,
\end{equation}
and analogously
\begin{equation}
    v_k
    =
    v_{\min}
    \left(
    \frac{v_{\max}}{v_{\min}}
    \right)^{\frac{k}{n_v-1}},
    \qquad
    k=0,\ldots,n_v-1.
\end{equation}
The intervals $[u_{\min},u_{\max}]$ and $[v_{\min},v_{\max}]$ determine the range of represented spatial scales, whereas $n_u$ and $n_v$ determine the frequency resolution.

We evaluate the spectrum at
\[
\mathcal U
=
\{0,\pm u_0,\ldots,\pm u_{n_u-1}\},
\]
and
\[
\mathcal V
=
\{0,v_0,\ldots,v_{n_v-1}\}.
\]
Only non-negative vertical frequencies are required because the contour measure is real-valued and hence its Fourier transform satisfies Hermitian symmetry,
\begin{equation}
    Z_{\mathcal A}(-u,-v)
    =
    Z_{\mathcal A}(u,v)^*.
\end{equation}

Unless otherwise stated, we use
\[
n_u=n_v=16,
\qquad
u_{\min}=v_{\min}=10^{-2},
\qquad
u_{\max}=v_{\max}=10,
\]
which results in
\[
(2n_u+1)(n_v+1)=33\times17=561
\]
sampled two-dimensional frequency pairs.

The one-dimensional camber and thickness spectra use the chordwise
frequency set
$\mathcal U_{\mathrm{1D}}=\{0,u_0,\ldots,u_{n_u-1}\}$.
Negative frequencies need not be stored separately because
$c_{\mathcal A}$ and $h_{\mathcal A}$ are real-valued and hence their
Fourier transforms satisfy Hermitian symmetry. In the implementation,
these one-dimensional spectra therefore share the chordwise parameters
$n_u$, $u_{\min}$, and $u_{\max}$ with the joint contour spectrum.

\subsection{Geometry Encoder Architectures}
\label{app:geometry-encoders}

The geometry-encoding ablation compares three representations of the
airfoil boundary
\[
\mathcal{A}=\{(x_i^a,y_i^a)\}_{i=1}^{N}.
\]
In all cases, the resulting geometry embedding
$g\in\mathbb{R}^{d_g}$ is subsequently supplied to the same downstream
surrogate together with the flow-field coordinate encoding and flow
conditions. The geometry encoders differ in how $\mathcal{A}$ is mapped
to $g$.

\paragraph{Raw-coordinate MLP baseline.}
The baseline operates directly on the sampled airfoil boundary points.
A shared pointwise MLP $\phi$ maps each boundary point to a hidden
representation,
\begin{equation}
    h_i=\phi(x_i^a,y_i^a),
\end{equation}
after which a permutation-invariant global pooling operation aggregates
the pointwise features,
\begin{equation}
    \bar h
    =
    \bigoplus_{i=1}^{N} h_i.
\end{equation}
A final learned projection $\rho$ produces the geometry embedding,
\begin{equation}
    g_{\mathrm{raw}}
    =
    \rho(\bar h).
\end{equation}
Thus, this encoder must learn the relevant global geometric information
directly from the sampled boundary coordinates.

\paragraph{Contour-level Poly2Vec.}
For Poly2Vec, the raw boundary is first mapped deterministically to its
two-dimensional contour spectrum
$Z_{\mathcal A}(u,v)$. After conversion of the complex coefficients to
real-valued features $\Phi(Z_{\mathcal A})$, a learned projection MLP
$P_Z$ produces a latent spectral representation,
\begin{equation}
    z=P_Z\!\left(\Phi(Z_{\mathcal A})\right).
\end{equation}
This representation is then passed through the common fusion network
$F$,
\begin{equation}
    g_{\mathrm{P2V}}
    =
    F([z,0,0]).
\end{equation}

\paragraph{Airfoil2Vec.}
Airfoil2Vec augments the joint contour spectrum with separate spectral
representations of the camber and half-thickness distributions,
denoted by $C_{\mathcal A}$ and $H_{\mathcal A}$, respectively. Three
learned projection networks produce
\begin{align}
    z &= P_Z\!\left(\Phi(Z_{\mathcal A})\right),\\
    c &= P_C\!\left(\Phi(C_{\mathcal A})\right),\\
    h &= P_H\!\left(\Phi(H_{\mathcal A})\right).
\end{align}
The final geometry embedding is
\begin{equation}
    g_{\mathrm{A2V}}
    =
    F([z,c,h]).
\end{equation}

\paragraph{Controlled Poly2Vec--Airfoil2Vec comparison.}
The Poly2Vec and Airfoil2Vec variants use the same projection dimensions
and the same fusion architecture. In implementation, all three spectral
branches are instantiated in both variants and the unused branches are
masked to zero. The two encoders can therefore be written jointly as
\begin{equation}
    g
    =
    F([
        m_Z z,\,
        m_C c,\,
        m_H h
    ]),
\end{equation}
where
\[
(m_Z,m_C,m_H)
=
(1,0,0)
\]
for contour-level Poly2Vec and
\[
(m_Z,m_C,m_H)
=
(1,1,1)
\]
for Airfoil2Vec.

Consequently, the Poly2Vec--Airfoil2Vec comparison does not change the
learned fusion architecture; it tests whether supplying the additional
airfoil-specific camber and thickness spectra improves the geometry
representation. The raw-coordinate baseline is architecturally different,
since it learns directly from boundary points through a shared pointwise
MLP and global pooling rather than from precomputed spectral features.

\subsection{Choosing the Fourier bandwidth and dimension.} 
\label{appendix:bandwith_dim}

The parameter $\sigma$ in the Gaussian Fourier projection controls the frequency range exposed to the network. We sweep $\sigma \in \{5, 15, 30, 45, 60, 75\}$ and the Fourier encoding dimension $l \in \{64, 128, 256, 512\}$. For low $\sigma$, the model is biased toward smooth functions and underfits sharper gradients. For high $\sigma$, optimization becomes harder and it is easier to overfit to noise. Table~\ref{tab:fourier_ablation} shows that $\sigma=45$ and $l=256$ give the best validation MSE among the tested settings. We therefore use $\sigma=45$ and $l=256$ in all following experiments.

\begin{table}[hbpt!]
    \centering
    \caption{Fourier ablation validation MSE (mean $\pm$ std) vs.\ $\sigma$ and encoding dimension.}
    \label{tab:fourier_ablation}
    \scalebox{0.6}{
    \begin{tabular}{@{} c c c c c c c | c c c c @{}}
        \toprule
        \multicolumn{7}{c|}{$\sigma$ sweep w/ $l=256$} & \multicolumn{4}{c}{$l$ sweep w/ $\sigma=45$} \\
        \cmidrule(lr){1-7} \cmidrule(lr){8-11}
         & 5 & 15 & 30 & 45 & 60 & 75 & 64 & 128 & 256 & 512 \\
        \midrule
        $(\times 10^{-4})$ & 2.16 & 2.00 & 2.16 & \textbf{1.77} & 2.08 & 2.08 & 1.96 & 1.89 & \textbf{1.77} & 2.31 \\
        & $\pm$0.21 & $\pm$0.06 & $\pm$1.08 & $\pm$\textbf{0.13} & $\pm$0.43 & $\pm$0.31 & $\pm$0.27 & $\pm$0.30 & $\pm$\textbf{0.13} & $\pm$0.60 \\
        \bottomrule
    \end{tabular}}
\end{table}

\section{Training Procedure}\label{app:training procedure}
\paragraph{Data splits and preprocessing.} The dataset consists of 9,977 flow-field simulations spanning 907 airfoil geometries, each evaluated at 11 angles of attack. We use the same 11 AoAs, $\alpha\in\{0^\circ,-1^\circ,\dots,-10^\circ\}$, for every geometry ($907\times11=9{,}977$). The flow fields are partitioned into a training set and a validation set that share no flow field (Table~\ref{tab:data_split}). The 47 geometries of the geometry interpolation and geometry extrapolation subsets are withheld at all 11 AoAs ($47\times11=517$ flow fields). For each of the remaining 860 geometries, the $\alpha=-5^\circ$ flow field is withheld to form the AoA interpolation subset (860 flow fields), and the other 10 AoAs form the training set ($860\times10=8{,}600$ flow fields). The validation set therefore contains $860+517=1{,}377$ flow fields. The AoA interpolation subset contains geometries that also appear in the training set (at different AoAs), whereas the geometry interpolation and extrapolation subsets contain only geometries never seen during training. The validation set is used for early stopping, checkpoint selection and model comparison (Section~\ref{sec:Quantitative Results}), and provides the cases analyzed in Section~\ref{sec:Aerodynamic analysis}; no separate test split is reserved. The five non-NACA airfoils of Appendix~\ref{app:non-NACA airfoils} are not part of the dataset and were used neither for training nor for model selection.

\begin{table}[h]
  \centering
  \caption{Composition of the training and validation sets. Every geometry is simulated at 11 AoAs, $\alpha\in\{0^\circ,-1^\circ,\dots,-10^\circ\}$. In the geometry extrapolation subset, $MP17$ and $MP18$ denote NACA$MP17$ and NACA$MP18$ for $M\in\{7,8\}$ and $P\in\{1,\dots,8\}$.}
  \label{tab:data_split}
  \small
  \begin{tabular}{l l p{4.4cm} c c}
    \toprule
    Set & Subset & Geometries & AoAs per geometry & Flow fields \\
    \midrule
    Training & -- & 860 NACA geometries & 10 (all but $-5^\circ$) & 8,600 \\
    \midrule
    \multirow{3}{*}{Validation} & AoA interpolation & the 860 training geometries & 1 ($-5^\circ$) & 860 \\
    & Geometry interpolation & 10: NACA1406, 2406, 4409, 4609, 6409, 6609, 1412, 2412, 4412, 6412 & 11 & 110 \\
    & Geometry extrapolation & 37: NACA8404, 8504, 8604, 8704, 8804; $MP17$, $MP18$ & 11 & 407 \\
    \midrule
    Total & & 907 & & 9,977 \\
    \bottomrule
  \end{tabular}
\end{table}

All input and output quantities are normalized to the $[0,1]$ range using min--max scaling computed on the training set. Specifically, the spatial coordinates $(x,y)$ and the output fields $(p,v_x,v_y)$ are each normalized independently per column. The AoA $\alpha$, which ranges from $-10^{\circ}$ to $0^{\circ}$, is rescaled by dividing by 10 to obtain values in $[-1,0]$. 

Each airfoil geometry is represented by a spline discretized into $N=2{,}000$ chordwise-aligned points (1,000 per surface), sampled with the sinusoidal chordwise distribution described in Section~\ref{sec:Spectral Geometry-Conditioning of Airfoils for Neural Surrogate Models} (spacing $\Delta x\in[10^{-6},1.6\times10^{-3}]c$, with points clustered near the leading and trailing edges). The resulting coordinates are stored in a lookup table and loaded into memory. The flow-field data are stored in the WebDataset format as sharded tar archives to enable efficient streaming during training. Because different meshes yield different cell counts, we employ a cell-sampling strategy: during each forward pass, $N_{\text{cells}} = 4{,}096$ cells are drawn uniformly at random from the full mesh of each flow field. This fixed-size sub-sampling ensures consistent batch dimensions while exposing the network to the entire spatial domain over the course of training.

\paragraph{Loss function.} The model is trained by minimizing the mean squared error (MSE) between predicted and ground-truth (normalized) quantities, summed across all three output fields:
\begin{equation}
    \mathcal{L} = \frac{1}{3N_{\text{cells}}}\sum_{j\in\{p,\,v_x,\,v_y\}}\sum_{i=1}^{N_{\text{cells}}} \left(\hat{q}_{j,i} - q_{j,i}\right)^2,
    \label{eq:loss_app}
\end{equation}
where $\hat{q}_{j,i}$ and $q_{j,i}$ denote the predicted and reference values of field $j$ at cell $i$, respectively. Per-head MSE and mean absolute error (MAE) are logged separately for each output quantity to monitor convergence and diagnose per-field behavior.

\paragraph{Optimization.} We use the AdamW optimizer with an initial learning rate of $10^{-3}$, weight decay $10^{-2}$, and default momentum parameters $(\beta_1, \beta_2) = (0.9, 0.999)$. The learning rate follows a cosine annealing schedule that decays from $10^{-3}$ to $10^{-6}$ over the full training horizon. Gradient norms are clipped to a maximum value of $1.0$ to stabilize training.

\paragraph{Training details.} All models are trained for a maximum of 100 epochs with a batch size of 64 flow fields (each sub-sampled to $4{,}096$ cells) using mixed-precision (BF16) arithmetic. Training is distributed across 8 GPUs with the PyTorch Distributed Data Parallel (DDP) strategy. We employ early stopping with a patience of 10 epochs on the validation loss (MSE over the full validation set), and save the top-3 checkpoints ranked by validation loss. All experiments are repeated over three random seeds to assess variance and report aggregate metrics.

\section{Angle of Attack and Geometry Interpolation Tests}\label{app:AoA-geometry Interpolation Tests}
\begin{figure}[hbpt!]
    \centering
    \includegraphics[width=0.8\linewidth]{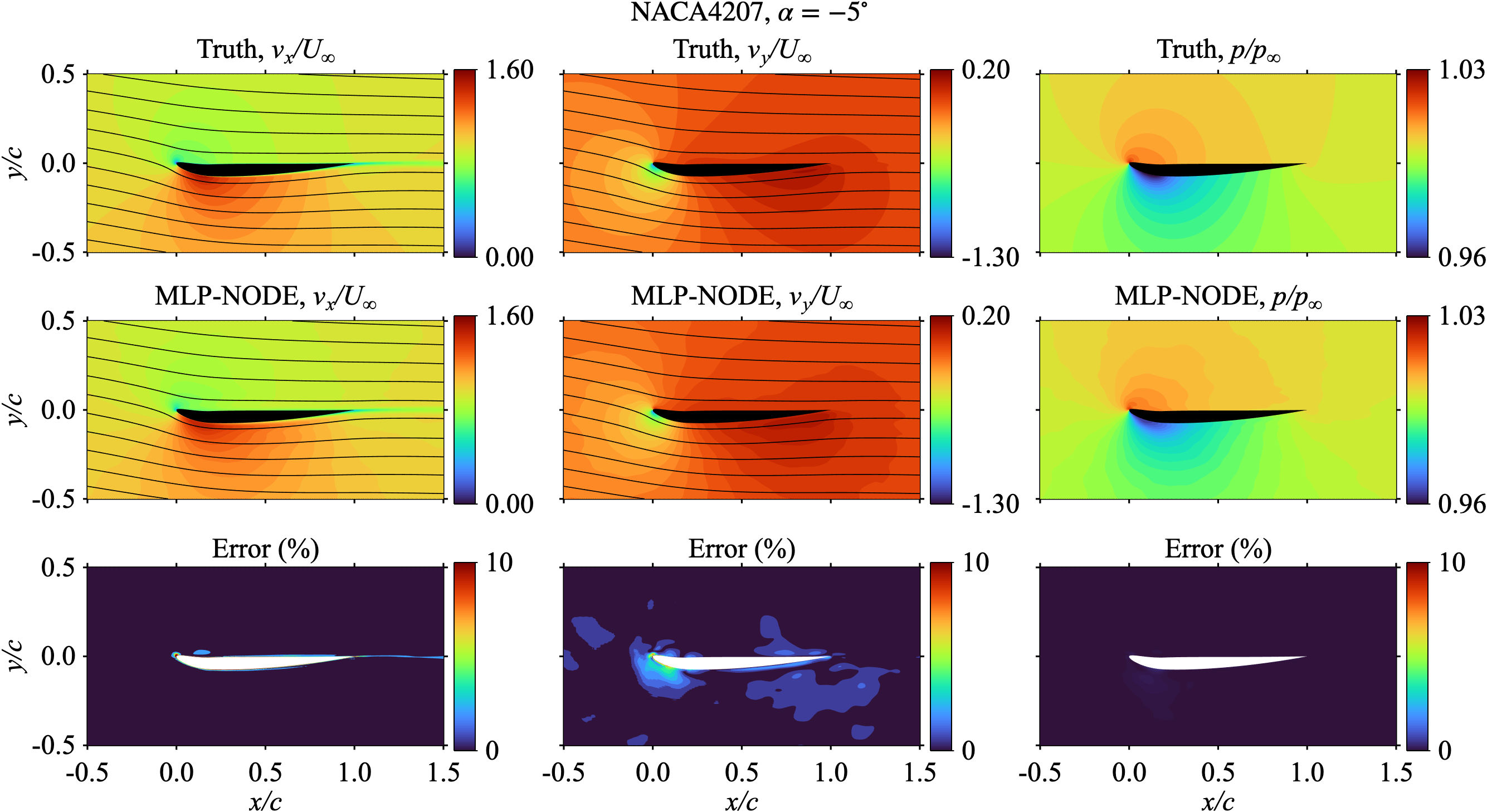}
    \caption{Comparison of (first column) $v_x/U_\infty$, (second column) $v_y/U_\infty$ and (third column) $p/p_\infty$ between OpenFOAM and the MLP-NODE prediction. Streamlines are superimposed. This case is representative of the capability of the model to interpolate AoA since the model was trained on the NACA4207 using all but $\alpha=-5^{\circ}$. Overall $\mathcal{L}_2$-norm based error $[\mathcal{L}_{2,v_x},\mathcal{L}_{2,v_y},\mathcal{L}_{2,p}] = [0.54,3.44,0.02]\%$.}
    \label{fig:aoa_interp}
\end{figure}

Figure~\ref{fig:aoa_interp} presents a comparison of $v_x/U_\infty$, $v_y/U_\infty$ and $p/p_\infty$ between the ground truth computed by OpenFOAM and the MLP-NODE model prediction for the NACA4207 at $\alpha = -5^{\circ}$. This case, taken from the AoA interpolation subset of the validation set, is representative of thin, medium-cambered airfoils. The NACA4207 at $\alpha = -5^{\circ}$ is chosen to showcase the ability of the model to interpolate new angles of attack, as the model was trained on the NACA4207 using all angles but $\alpha = -5^{\circ}$. The figure shows that the model predictions are in excellent agreement with the CFD data. The pressure field $\mathcal{L}_2$-norm error remains well below 1\%, while the $v_x$ and $v_y$ velocity components exhibit slightly higher errors, particularly in the near-wall, nose and wake regions. Nevertheless, the pointwise percentage error remains below $10\%$ over most of the field of view and $<4\%$ in an $\mathcal{L}_2$-norm sense, indicating robust predictive performance overall.

\begin{figure}[hbpt!]
    \centering
    \includegraphics[width=0.8\linewidth]{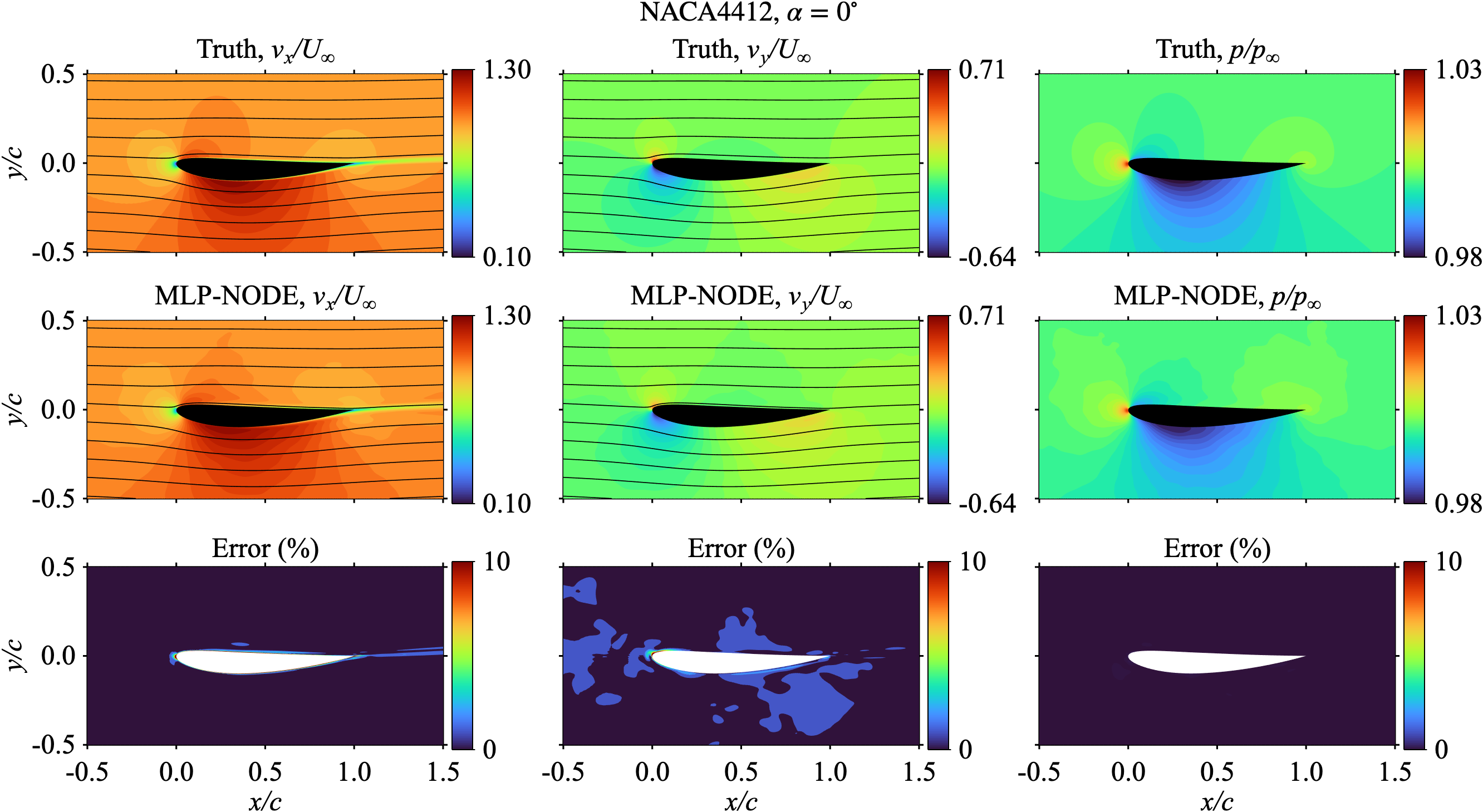}
    \caption{Comparison of (first column) $v_x/U_\infty$, (second column) $v_y/U_\infty$ and (third column) $p/p_\infty$ between OpenFOAM and the MLP-NODE prediction. This case is representative of the model’s ability to interpolate airfoil geometry, since the model was trained on geometries similar to the NACA4412 at $\alpha = 0^{\circ}$, as shown. Overall $\mathcal{L}_2$-norm based error $[\mathcal{L}_{2,v_x},\mathcal{L}_{2,v_y},\mathcal{L}_{2,p}] = [0.53, 8.34, 0.01]\%$.}
    \label{fig:geom_interp}
\end{figure}

Next, Figure~\ref{fig:geom_interp} compares the reference and MLP-NODE predicted data for the NACA4412 airfoil. This geometry, from the geometry interpolation subset of the validation set, is selected to assess the model’s ability to interpolate airfoil shapes, as it was not included in the training set at any AoA but lies within the parametric bounds of the training data, with neighboring geometries above and below it in camber and thickness. The magnitudes and spatial trends are comparable to those observed for the previous case in Figure~\ref{fig:aoa_interp}, indicating robust performance in this scenario as well.

\section{Interpolation and Extrapolation of the Pressure Coefficient Around the Airfoil}\label{app: Pressure coefficient interpolation-extrapolation}
\begin{figure}[hbpt!]
    \centering
    \includegraphics[width=\linewidth]{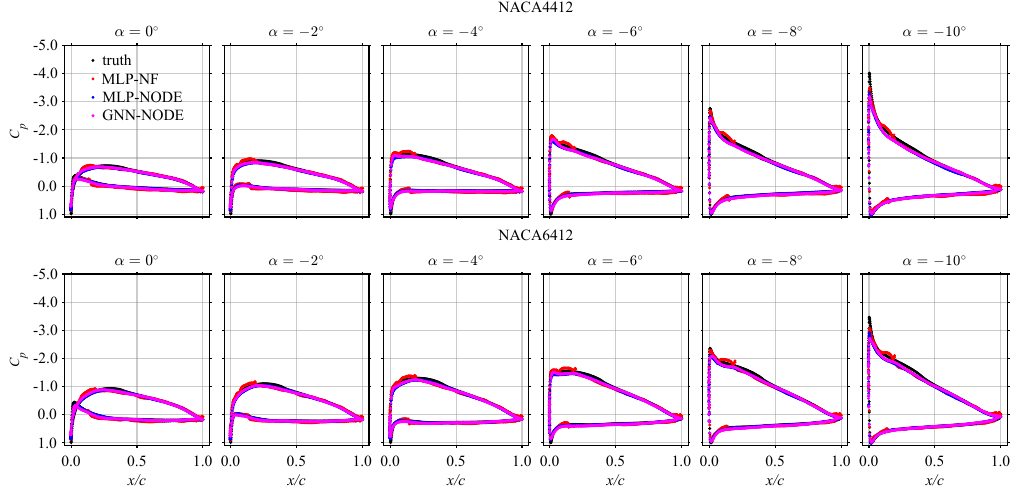}
    \caption{Pressure coefficient at several $\alpha$ for two airfoil geometries featuring medium values of camber and thickness. This demonstrates geometry interpolation capability of the models: both airfoils belong to the geometry interpolation subset of the validation set and were not seen during training at any AoA.}
    \label{fig:Cp_geom_interp}
\end{figure}

Figure~\ref{fig:Cp_geom_interp} shows the pressure coefficient distributions of two NACA airfoils tested for geometry interpolation across the angle of attack polar. Similar trends to Figure~\ref{fig:Cp_AoA_interp} are observed; the MLP-NF model matches the magnitude of the suction peak more closely than the NODE models but lacks smoothness. Overall, the NODE models' predictions are almost overlapping and capture the fundamental $C_p$ curve shape at all incidences. As the magnitude of the negative AoA increases, sharper suction peaks are produced by the flow and the spread between the models becomes more visible because of the smoothening effect of the NODE models.

\begin{figure}[hbpt!]
    \centering
    \includegraphics[width=.85\linewidth]{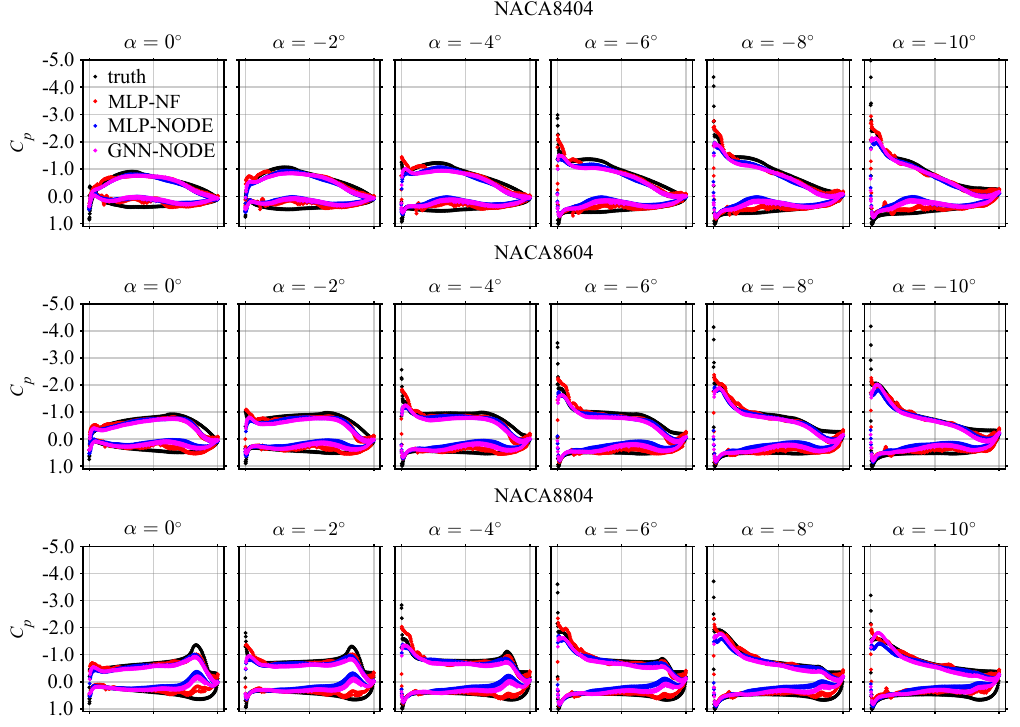} \\
    \includegraphics[width=.85\linewidth]{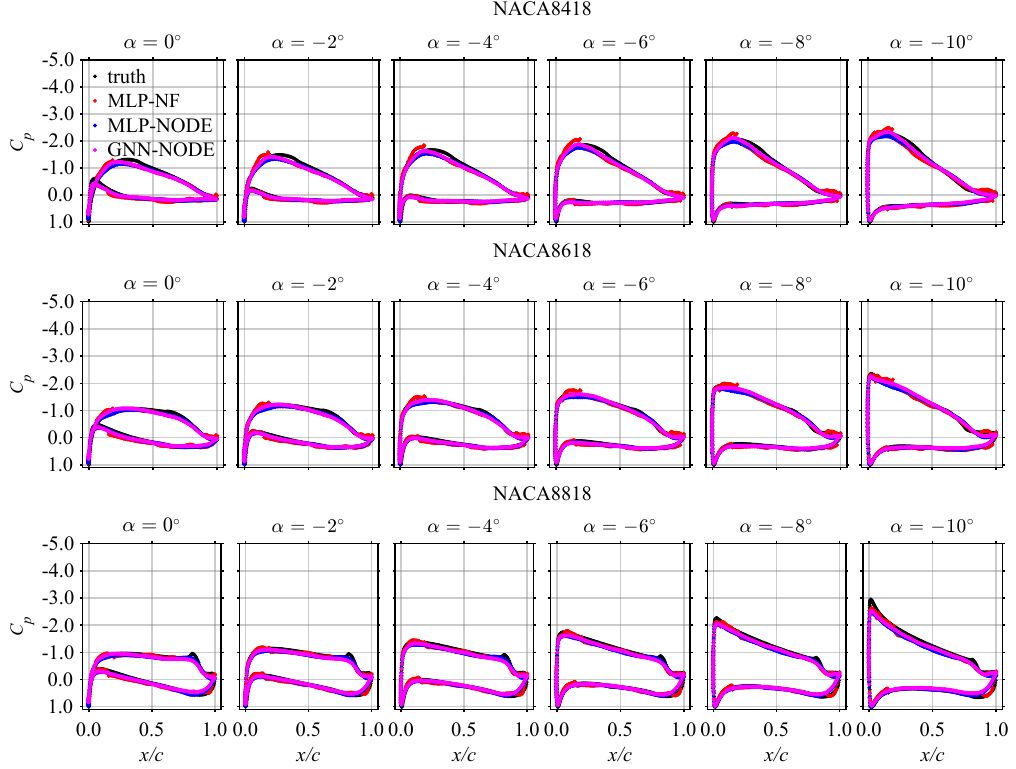}
    \caption{Pressure coefficient at several $\alpha$ for a few airfoils with high camber (8\%), variable position of maximum camber and very thin (4\%) and thick (18\%) geometric features. This demonstrates geometry extrapolation capability of the models: all airfoils shown belong to the geometry extrapolation subset of the validation set and were not seen during training at any AoA.}
    \label{fig:Cp_geom_extrap}
\end{figure}

From a geometry extrapolation standpoint, the accuracy of the models' prediction is different between thin and thick (cambered) airfoils --see Figure~\ref{fig:Cp_geom_extrap}. Among thin airfoils, the MLP-NF model solution is strongly oscillatory. The NODE models are more accurate up to the mid-chord, but the rear pressure loading shows discrepancies with respect to the CFD data. In particular, aft-cambered geometries produce suction in the rear surface of the airfoil introducing a distinctive peak of negative $C_p$. All models struggle with resolving this peak; the location is accurately identified but the magnitude is underestimated. The pressure rise associated with flow deceleration on the upper surface is also not properly resolved, since the models predict suction (local flow acceleration) there. This is visible from non-physical local peaks of (negative) $C_p$ in the lower part of the $C_p$ shape in NACA8804. On the other hand, thick airfoils are matched well by the models. The suction in the lower surface of the NACA8818 airfoil (upper $C_p$ branch) is either missed or heavily damped by model predictions, however the lower $C_p$ branch (upper airfoil surface) follows the CFD solution closely.

\section{Prediction of the Wake Region}\label{app:wake prediction}
Tables~\ref{tab:wake_l2_error_naca_4_thickness} and~\ref{tab:wake_l2_error_naca_18_thickness} complement figure~\ref{fig:wake_airfoil} by quantifying the $\mathcal{L}_2$-norm error of the reconstructed $v_x/U_\infty$ field in the wake patch $x/c\in[1,2]$, $y/c\in[-0.2,0.2]$ for the same thin and thick airfoil sets, for the three models (MLP-NF, MLP-NODE, GNN-NODE) and for the full angle-of-attack polar $\alpha\in[-10^\circ,0^\circ]$. The most accurate model at each condition is highlighted in green, and the mean, standard deviation, extrema and coefficient of variation (CoV) over the polar are reported for each column. Four main observations follow.

\paragraph{Position of maximum camber.} The under-prediction of the velocity deficit observed for aft-cambered sections in figure~\ref{fig:wake_airfoil} translates into a monotonic growth of the polar-averaged error as the maximum camber moves from $4/10$ to $8/10$ of the chord: the mean error of MLP-NODE rises from $1.64\%$ to $4.07\%$ in the thin set and from $1.96\%$ to $4.25\%$ in the thick set, and the same ordering holds for MLP-NF and GNN-NODE. The effect is not confined to high loading: at $\alpha=0^\circ$ the error of the aft-loaded sections is already two to three times that of the forward-loaded ones. The camber position therefore sets the baseline level of the wake error, on top of which the incidence acts.

\paragraph{Angle of attack.} In the thick set the error increases smoothly and almost monotonically with $|\alpha|$ for every geometry and model, by no more than $\sim\!4$ percentage points between $\alpha=0^\circ$ and $\alpha=-10^\circ$. The thin set is less regular: the forward-camber sections show a shallow minimum at $\alpha=-1^\circ$ to $-3^\circ$ rather than at zero incidence, GNN-NODE is non-monotonic between $\alpha=-7^\circ$ and $-9^\circ$, and, most notably, the aft-most section NACA8804 exhibits an abrupt jump at $\alpha=-10^\circ$, where the error grows with respect to $\alpha=-9^\circ$ by factors of $1.9$, $1.5$ and $2.0$ for MLP-NF, MLP-NODE and GNN-NODE, respectively (from $4.8$--$5.3\%$ to $7.0$--$10.8\%$) and reaches the largest values of either table. As $\alpha=-10^\circ$ is the edge of the training polar, this jump is more plausibly an extrapolation effect on a strongly deflected thin-airfoil wake than a smooth loss of accuracy with loading.

\begin{table}[hbpt!]
    \centering
    \caption{Wake $v_x/U_\infty$ $\mathcal{L}_2$-norm error (\%) in the field of view $x/c\in[1,2]$, $y/c\in[-0.2,0.2]$ for models MLP-NF, MLP-NODE, GNN-NODE on NACA airfoils with 8\% camber and 4\% thickness, representative of thin cambered airfoils.}
    \label{tab:wake_l2_error_naca_4_thickness}
    \resizebox{\textwidth}{!}{%
        \begin{tabular}{c c c c c c c c c c c c c c c c}
        \toprule
        & \multicolumn{3}{c}{NACA8404} & \multicolumn{3}{c}{NACA8504} & \multicolumn{3}{c}{NACA8604} & \multicolumn{3}{c}{NACA8704} & \multicolumn{3}{c}{NACA8804} \\
        \cmidrule(lr){2-4} \cmidrule(lr){5-7} \cmidrule(lr){8-10} \cmidrule(lr){11-13} \cmidrule(lr){14-16}
        $\alpha$ $[^\circ]$ & NF & MLP & GNN & NF & MLP & GNN & NF & MLP & GNN & NF & MLP & GNN & NF & MLP & GNN \\
        \midrule
        0 & 1.53 & \textcolor{green}{1.24} & 1.35 & 1.60 & 1.16 & \textcolor{green}{1.13} & 1.82 & \textcolor{green}{1.42} & 1.43 & 2.23 & \textcolor{green}{1.84} & 2.14 & 2.77 & 2.65 & \textcolor{green}{2.31} \\
        -1 & 1.38 & \textcolor{green}{1.00} & \textcolor{green}{1.00} & 1.50 & 1.10 & \textcolor{green}{1.04} & 1.73 & \textcolor{green}{1.40} & \textcolor{green}{1.40} & 2.14 & \textcolor{green}{1.71} & 1.94 & 2.58 & 2.71 & \textcolor{green}{2.44} \\
        -2 & 1.32 & \textcolor{green}{1.00} & 1.02 & 1.39 & 1.11 & \textcolor{green}{1.05} & 1.59 & 1.39 & \textcolor{green}{1.34} & 2.06 & \textcolor{green}{1.67} & 1.83 & \textcolor{green}{2.60} & 2.97 & 2.90 \\
        -3 & 1.24 & \textcolor{green}{1.01} & 1.03 & 1.30 & 1.33 & \textcolor{green}{1.20} & \textcolor{green}{1.49} & 1.56 & 1.50 & 2.25 & \textcolor{green}{2.12} & 2.34 & \textcolor{green}{3.10} & 3.63 & 3.79 \\
        -4 & 1.22 & 1.20 & \textcolor{green}{1.18} & \textcolor{green}{1.31} & 1.67 & 1.53 & \textcolor{green}{1.66} & 2.01 & 2.01 & 2.80 & \textcolor{green}{2.68} & 3.03 & \textcolor{green}{3.60} & 3.95 & 4.26 \\
        -5 & \textcolor{green}{1.32} & 1.39 & 1.75 & \textcolor{green}{1.56} & 2.06 & 2.18 & \textcolor{green}{2.27} & 2.38 & 2.62 & 3.41 & \textcolor{green}{2.90} & 3.37 & 3.93 & \textcolor{green}{3.65} & 4.15 \\
        -6 & 1.73 & \textcolor{green}{1.65} & 2.75 & \textcolor{green}{2.14} & 2.29 & 2.98 & 3.07 & \textcolor{green}{2.58} & 3.09 & 3.88 & \textcolor{green}{2.84} & 3.35 & 4.42 & \textcolor{green}{3.92} & 4.51 \\
        -7 & 2.40 & \textcolor{green}{1.90} & 3.61 & 2.70 & \textcolor{green}{2.37} & 3.39 & 3.58 & \textcolor{green}{2.64} & 3.12 & 4.31 & \textcolor{green}{3.05} & 3.38 & 5.04 & \textcolor{green}{4.58} & 4.61 \\
        -8 & 2.68 & \textcolor{green}{1.95} & 3.58 & 2.79 & \textcolor{green}{2.48} & 2.92 & 3.73 & \textcolor{green}{2.71} & 2.75 & 4.47 & 3.12 & \textcolor{green}{3.08} & 4.68 & 4.89 & \textcolor{green}{4.54} \\
        -9 & 2.57 & \textcolor{green}{2.11} & 2.43 & 3.16 & 3.02 & \textcolor{green}{2.08} & 3.91 & 2.92 & \textcolor{green}{2.33} & 4.65 & \textcolor{green}{3.07} & 3.08 & 4.92 & \textcolor{green}{4.80} & 5.25 \\
        -10 & 3.94 & 3.64 & \textcolor{green}{3.05} & 4.86 & 4.70 & \textcolor{green}{3.67} & 4.94 & 4.43 & \textcolor{green}{4.36} & 5.03 & \textcolor{green}{3.43} & 5.26 & 9.13 & \textcolor{green}{7.00} & 10.76 \\
        \midrule
        mean & 1.94 & \textcolor{green}{1.64} & 2.07 & 2.21 & 2.12 & \textcolor{green}{2.11} & 2.71 & \textcolor{green}{2.31} & 2.36 & 3.38 & \textcolor{green}{2.58} & 2.98 & 4.25 & \textcolor{green}{4.07} & 4.50 \\
        std & 0.86 & 0.77 & 1.05 & 1.10 & 1.07 & 1.00 & 1.19 & 0.91 & 0.95 & 1.13 & 0.63 & 0.96 & 1.86 & 1.25 & 2.28 \\
        min & 1.22 & 1.00 & 1.00 & 1.30 & 1.10 & 1.04 & 1.49 & 1.39 & 1.34 & 2.06 & 1.67 & 1.83 & 2.58 & 2.65 & 2.31 \\
        max & 3.94 & 3.64 & 3.61 & 4.86 & 4.70 & 3.67 & 4.94 & 4.43 & 4.36 & 5.03 & 3.43 & 5.26 & 9.13 & 7.00 & 10.76 \\
        CoV\,(\%) & 44.6 & 47.1 & 50.6 & 49.8 & 50.6 & 47.3 & 44.0 & 39.4 & 40.2 & 33.5 & 24.4 & 32.1 & 43.7 & 30.6 & 50.6 \\
        \bottomrule
        \end{tabular}
    }
\end{table}

\paragraph{Thickness.} Consistent with the misprediction of the wake shape at $x/c\geq1.8$ in the thick set (figure~\ref{fig:wake_airfoil}), the polar-averaged error is higher for the thick than for the thin section in 13 of the 15 combinations of camber position and model; the exceptions are GNN-NODE on the 84-series, where both sections give $2.07\%$, and GNN-NODE on the 88-series, where NACA8818 ($4.41\%$) is slightly below NACA8804 ($4.50\%$). Thickness, however, makes the error far more predictable across the polar: the peak error drops from $10.76\%$ (thin, GNN-NODE on NACA8804) to $5.81\%$ (thick, GNN-NODE on NACA8818), and for the aft-loaded sections the standard deviation falls below $0.9\%$ and the CoV to $13$--$19\%$, against $24$--$51\%$ for their thin counterparts. Thickness thus trades a moderate increase of the mean wake error for a marked gain in robustness, whereas the thin set offers the best-case accuracy but also the least predictable extremes.

\begin{table}[htbp!]
    \centering
    \caption{Wake $v_x/U_\infty$ $\mathcal{L}_2$-norm error (\%) in the field of view $x/c\in[1,2]$, $y/c\in[-0.2,0.2]$ for models MLP-NF, MLP-NODE, GNN-NODE on NACA airfoils with 8\% camber and 18\% thickness, representative of thick cambered airfoils.}
    \label{tab:wake_l2_error_naca_18_thickness}
    \resizebox{\textwidth}{!}{%
        \begin{tabular}{c c c c c c c c c c c c c c c c}
        \toprule
        & \multicolumn{3}{c}{NACA8418} & \multicolumn{3}{c}{NACA8518} & \multicolumn{3}{c}{NACA8618} & \multicolumn{3}{c}{NACA8718} & \multicolumn{3}{c}{NACA8818} \\
        \cmidrule(lr){2-4} \cmidrule(lr){5-7} \cmidrule(lr){8-10} \cmidrule(lr){11-13} \cmidrule(lr){14-16}
        $\alpha$ $[^\circ]$ & NF & MLP & GNN & NF & MLP & GNN & NF & MLP & GNN & NF & MLP & GNN & NF & MLP & GNN \\
        \midrule
        0 & 1.69 & \textcolor{green}{0.88} & 1.68 & 1.63 & \textcolor{green}{0.93} & 1.62 & 1.81 & \textcolor{green}{1.32} & 1.87 & 3.01 & 2.41 & \textcolor{green}{2.30} & 3.71 & \textcolor{green}{3.25} & 3.39 \\
        -1 & 1.77 & \textcolor{green}{1.03} & 1.46 & 1.71 & \textcolor{green}{1.11} & 1.38 & 1.96 & \textcolor{green}{1.56} & 1.80 & 3.17 & 2.68 & \textcolor{green}{2.52} & 3.68 & 3.27 & \textcolor{green}{3.24} \\
        -2 & 1.87 & \textcolor{green}{1.22} & 1.38 & 1.82 & \textcolor{green}{1.32} & 1.34 & 2.20 & \textcolor{green}{1.88} & 2.00 & 3.43 & 3.01 & \textcolor{green}{2.91} & 3.87 & \textcolor{green}{3.59} & 3.64 \\
        -3 & 1.98 & \textcolor{green}{1.39} & 1.45 & 1.97 & 1.53 & \textcolor{green}{1.48} & 2.50 & \textcolor{green}{2.22} & 2.37 & 3.64 & \textcolor{green}{3.27} & 3.28 & 4.02 & \textcolor{green}{3.79} & 3.99 \\
        -4 & 2.12 & \textcolor{green}{1.54} & 1.59 & 2.15 & 1.74 & \textcolor{green}{1.71} & 2.72 & \textcolor{green}{2.41} & 2.67 & 3.75 & \textcolor{green}{3.41} & 3.50 & 4.16 & \textcolor{green}{4.01} & 4.18 \\
        -5 & 2.31 & \textcolor{green}{1.72} & 1.77 & 2.35 & \textcolor{green}{1.90} & 1.91 & 2.97 & \textcolor{green}{2.59} & 2.92 & 3.84 & \textcolor{green}{3.49} & 3.61 & 4.43 & \textcolor{green}{4.30} & \textcolor{green}{4.30} \\
        -6 & 2.56 & \textcolor{green}{1.92} & 1.97 & 2.61 & \textcolor{green}{2.09} & \textcolor{green}{2.09} & 3.27 & \textcolor{green}{2.78} & 3.11 & 3.95 & \textcolor{green}{3.59} & 3.64 & 4.25 & \textcolor{green}{4.17} & 4.40 \\
        -7 & 2.84 & \textcolor{green}{2.16} & 2.17 & 2.99 & 2.39 & \textcolor{green}{2.32} & 3.66 & \textcolor{green}{3.09} & 3.33 & 4.12 & 3.80 & \textcolor{green}{3.69} & \textcolor{green}{4.68} & \textcolor{green}{4.68} & 4.87 \\
        -8 & 3.24 & 2.53 & \textcolor{green}{2.45} & 3.52 & 2.87 & \textcolor{green}{2.70} & 4.15 & \textcolor{green}{3.56} & 3.65 & 4.36 & 4.15 & \textcolor{green}{3.83} & 4.87 & \textcolor{green}{4.76} & 5.12 \\
        -9 & 3.85 & 3.15 & \textcolor{green}{2.97} & 4.22 & 3.59 & \textcolor{green}{3.32} & 4.76 & 4.23 & \textcolor{green}{4.18} & 4.44 & 4.39 & \textcolor{green}{4.01} & \textcolor{green}{5.14} & 5.22 & 5.54 \\
        -10 & 4.69 & 4.05 & \textcolor{green}{3.84} & 5.10 & 4.55 & \textcolor{green}{4.28} & 5.46 & 5.09 & \textcolor{green}{4.96} & 4.55 & 4.61 & \textcolor{green}{4.29} & \textcolor{green}{5.52} & 5.66 & 5.81 \\
        \midrule
        mean & 2.63 & \textcolor{green}{1.96} & 2.07 & 2.73 & \textcolor{green}{2.18} & 2.20 & 3.22 & \textcolor{green}{2.79} & 2.99 & 3.84 & 3.53 & \textcolor{green}{3.42} & 4.39 & \textcolor{green}{4.25} & 4.41 \\
        std & 0.95 & 0.97 & 0.76 & 1.13 & 1.11 & 0.92 & 1.18 & 1.14 & 1.00 & 0.51 & 0.69 & 0.61 & 0.60 & 0.78 & 0.85 \\
        min & 1.69 & 0.88 & 1.38 & 1.63 & 0.93 & 1.34 & 1.81 & 1.32 & 1.80 & 3.01 & 2.41 & 2.30 & 3.68 & 3.25 & 3.24 \\
        max & 4.69 & 4.05 & 3.84 & 5.10 & 4.55 & 4.28 & 5.46 & 5.09 & 4.96 & 4.55 & 4.61 & 4.29 & 5.52 & 5.66 & 5.81 \\
        CoV\,(\%) & 36.3 & 49.1 & 36.9 & 41.2 & 50.8 & 41.8 & 36.6 & 40.9 & 33.5 & 13.2 & 19.4 & 18.0 & 13.6 & 18.3 & 19.2 \\
        \bottomrule
        \end{tabular}
    }
\end{table}

\paragraph{Model comparison.} MLP-NODE attains the lowest polar-averaged error on eight of the ten geometries and is within $0.11$ percentage points of GNN-NODE on the remaining two (NACA8504: $2.12\%$ vs.\ $2.11\%$; NACA8718: $3.53\%$ vs.\ $3.42\%$); MLP-NF is never the best on average. The advantage of the NODE models is concentrated at low incidence, where their error is up to about half that of MLP-NF, while MLP-NF becomes competitive or marginally the best at intermediate incidence ($\alpha=-3^\circ$ to $-6^\circ$) in the thin set and at the highest incidences on NACA8818. GNN-NODE is the most accurate model at $\alpha=-10^\circ$ for all forward-camber sections (NACA84xx--86xx) of both sets, and at $\alpha\leq-8^\circ$ for NACA8418--8718 in the thick set, with the single exception of NACA8618 at $\alpha=-8^\circ$, where MLP-NODE is marginally better ($3.56\%$ vs.\ $3.65\%$), but it is also the least robust on the thin aft-loaded sections, where it produces the largest error and CoV of either table. MLP-NODE combines the lowest mean with the lowest peak error in the thin set and a peak within $0.15$ points of the best in the thick set, which motivates its selection for the wake profiles of figure~\ref{fig:wake_airfoil}.

In summary, the wake error is governed primarily by the position of maximum camber and secondarily by the incidence, with thickness acting as a regularizer that raises the mean error slightly but suppresses the extreme errors and the irregular incidence dependence seen in thin, aft-loaded sections.

\section{Spectral Noise analysis}\label{sec:Spectral Noise analysis}
Let $q(\mathbf{x})$ denote a normalized state ($v_x/U_\infty$, $v_y/U_\infty$ or $p/p_\infty$), with $q_t$ the CFD truth and $q_p$ a model prediction, both available at the $N$ cell centers of the computational mesh. Spatially uncorrelated small-scale error (``noise'') is diagnosed in the wavenumber domain. Both fields are mapped with the same piecewise-linear interpolant $\mathcal{I}$ on the Delaunay triangulation of the cell centers \cite{amidror2002scattered} onto a uniform $n_g\times n_g$ grid of spacing $\Delta=L/(n_g-1)$ covering a square probe patch $\Omega$ of side $L$. Two patches are used (Table~\ref{tab:psd_settings}): a farfield patch in clean flow above the airfoil and a near-wake patch downstream of the trailing-edge, where the flow is populated by the wake shear layer. Linear interpolation preserves the point-to-point roughness of the data -- its error for a smooth field is $O(h^2\,|\nabla^2 q|)$, $h$ being the local cell size \cite{ciarlet1978fem} -- so mesh-scale noise in $q_p$ survives the gridding provided $\Delta\le h_{\min}/2$. This holds in the farfield patch; in the wake patch only the few trailing-edge cells ($h\approx0.002c$) are sampled at $\Delta\approx h$, and their aliased imprint is negligible.

Each gridded field is demeaned over $\Omega$, multiplied by a 2-D Hann window $w$ \cite{harris1978windows} and transformed with the 2-D DFT. The periodogram \cite{welch1967psd} is normalized so that Parseval's identity holds \cite{bracewell2000fourier}:
\begin{equation}
  P(\mathbf{k})=\frac{\big|\mathcal{F}\{(q-\bar q)\,w\}(\mathbf{k})\big|^2}{D},\qquad D=n_g^2\sum_{\Omega}w^2,
  \qquad \sum_{\mathbf{k}}P(\mathbf{k})=\operatorname{Var}_w(q).
\end{equation}
Isotropic spectra follow by integrating $P$ over annular shells of the physical wavenumber $\kappa=|\mathbf{k}|/L$ (cycles per chord) of width $\Delta\kappa=1/L$ \cite{Pope2000Turbulent}:
\begin{equation}
  E_t(\kappa)=\sum_{\text{shell}}P_t(\mathbf{k}),\qquad
  E_p(\kappa)=\sum_{\text{shell}}P_p(\mathbf{k}),\qquad
  E_{p-t}(\kappa)=\frac{1}{D}\sum_{\text{shell}}|F_p-F_t|^2,
\end{equation}
where $F_t=\mathcal{F}\{(q_t-\bar q_t)\,w\}$ and $F_p=\mathcal{F}\{(q_p-\bar q_p)\,w\}$ are the windowed transforms and $E_{p-t}$ is the spectrum of the residual $q_p-q_t$ (the DFT being linear). $E$ is a shell-integrated energy with the units of $q^2$: it sums to the windowed variance, and an uncorrelated (white) field appears as $E\propto\kappa$ because the number of modes per shell grows linearly with $\kappa$.

Three diagnostics are formed per shell. The energy ratio $E_p/E_t$ isolates energy the model places at scales where the truth has none ($E_p/E_t\gg1$: added or amplified content; $<1$: attenuated content, i.e. over-smoothing), the two behaviors expected from the spectral bias of coordinate networks \cite{rahaman2019spectral,tancik2020fourier}. The spectral coherence \cite{bendat2010random}
\begin{equation}
  \gamma^2(\kappa)=\frac{\big|\sum_{\text{shell}}F_p\,\overline{F_t}\big|^2}
  {\big(\sum_{\text{shell}}|F_p|^2\big)\big(\sum_{\text{shell}}|F_t|^2\big)}\in[0,1]
\end{equation}
is the fraction of the prediction energy at scale $1/\kappa$ that is linearly related to the truth; $(1-\gamma^2)E_p$ is its incoherent part. The cumulative tail RMS
\begin{equation}
  \sigma_\kappa(\kappa)=\Big(\sum_{\kappa'\ge\kappa}E_{p-t}(\kappa')\Big)^{1/2}
\end{equation}
is the RMS of the residual band-limited to wavelengths shorter than $1/\kappa$. It is a high-frequency residual metric rather than a noise measure per se: the scalar reported is the high-frequency residual $\sigma_{\text{hf}}=\sigma_\kappa(\kappa_c)$, and the conditions under which it can be read as added noise are stated below. By linearity of the DFT,
\begin{equation}
  E_{p-t}(\kappa)=E_p(\kappa)+E_t(\kappa)-\frac{2}{D}\,\Re\sum_{\text{shell}}F_p\overline{F_t},
\end{equation}
so $\sigma_\kappa$ reduces to $(\sum_{\kappa'\ge\kappa}E_p)^{1/2}$ wherever the truth is empty -- the regime realized in the farfield patch. $E_{p-t}$ is always evaluated directly from its definition; this identity serves interpretation only. In the wake patch the truth is not empty above the cutoff and the two diagnostics separate the residual: a model that over-smooths has $E_p/E_t\ll1$ and its $\sigma_\kappa$ collapses onto the truth tail $(\sum_{\kappa'\ge\kappa}E_t)^{1/2}$, i.e. missing content rather than noise, whereas a model that adds noise has $E_p/E_t\gg1$, low coherence and $\sigma_\kappa$ above the truth tail.

\paragraph{Cutoff wavenumber.} $\kappa_c$ is set from the truth alone (truth-tail criterion) as the smallest shell above which the truth retains less than a tolerance $\varepsilon$ of its variance,
\begin{equation}
  \kappa_c=\min\Big\{\kappa:\ \sum_{\kappa'\ge\kappa}E_t(\kappa')\le\varepsilon\sum_{\kappa'}E_t(\kappa')\Big\},\qquad \varepsilon=10^{-3},
\end{equation}
so that the band $[\kappa_c,\kappa_{\max}]$ holds a negligible share of the physical content, at most $\varepsilon\operatorname{Var}_w q_t$. The admissible band is bounded by the discretization scales: the patch size fixes the fundamental $\Delta\kappa=1/L$ and, with a Hann window, a leakage floor of about four fundamentals \cite{harris1978windows}; the grid fixes the Nyquist limit $\kappa_{\max}=(n_g-1)/(2L)$ \cite{shannon1949}; the mesh fixes the finest resolvable physics, $\kappa_{\text{mesh}}=1/(2h)$, and the scale of the interpolation imprint, $\kappa\approx1/h$. A conservative cutoff is therefore
\begin{equation}
  \kappa_c^{*}=\max\!\big(\kappa_c,\,4/L\big),\qquad
  \frac{4}{L}\le\kappa_c^{*}<\frac{1}{2h_{\max}}<\kappa_{\max}=\frac{n_g-1}{2L}.
\end{equation}
Table~\ref{tab:psd_settings} lists the settings and the derived scales for both patches for NACA8604 at $\alpha=-5^\circ$ and $q=v_y/U_\infty$. In the farfield patch the truth spectrum decays so fast that the criterion is bound by the leakage floor, $\kappa_c=4/L=4$ cycles/$c$, and the whole band up to $\kappa_{\max}$ measures content added by the model. In the wake patch the criterion is bound by the physics, $\kappa_c=18$ cycles/$c$: the shear layer populates wavelengths down to $\approx0.06c$, and the cutoff still satisfies $\kappa_c<1/(2h_{\max})=34$ cycles/$c$. There the truth tail provides a reference level for over-smoothing, the omission bound $B=(\varepsilon\operatorname{Var}_w q_t)^{1/2}$: by construction of $\kappa_c$, the truth-tail energy $T=\sum_{\kappa\ge\kappa_c}E_t$ satisfies $T\le B^2$, so a prediction that only attenuates or omits the truth tail has $\sigma_{\text{hf}}\le\sqrt{T}\le B$, the first equality holding when it omits the tail entirely. $B$ is not an upper bound for every noise-free prediction, as the controlled cases below show. A patch containing the airfoil is not admissible: zeroing the body interior creates a discontinuity of transverse width equal to the airfoil thickness $t$, whose transform $t\,\operatorname{sinc}(t\kappa_y)$ \cite{bracewell2000fourier} rings with lobes spaced $1/t$ across the whole band and dominates $E_t$ up to the Nyquist limit; on a full-domain grid with $\Delta=0.01c$ the criterion is then pushed to $\kappa_c\approx25$-$45$ cycles/$c$ against $\kappa_{\max}=50$.

\paragraph{Controlled cases.} The diagnostics respond differently to the elementary ways in which a prediction can depart from the truth above $\kappa_c$. Table~\ref{tab:controlled_cases} lists, directly from the definitions above, their response to five cases: an exact prediction; attenuation, $F_p=aF_t$ with $0\le a\le1$ ($a=0$ being omission, i.e.\ complete over-smoothing); coherent amplification, $F_p=aF_t$ with $a>1$; a spatial displacement $\mathbf s$ of the truth, idealized as the phase shift $F_p=F_t\,e^{-2\pi\mathrm{i}\,\mathbf k\cdot\mathbf s/L}$; and additive noise $n$ uncorrelated with the truth, $F_p=F_t+F_n$. Two conclusions follow. First, $\sigma_{\text{hf}}>B$ does not by itself indicate noise: a perfectly coherent amplification ($\gamma^2=1$, $\sigma_{\text{inc}}=0$) exceeds $B$ once $a>1+B/\sqrt{T}\ge2$; for instance, $F_p=3F_t$ gives $\sigma_{\text{hf}}=2\sqrt{T}$, which exceeds $B$ whenever $T>B^2/4$. Second, a residual can be attributed to added noise only when (a)~$E_p/E_t$ substantially exceeds unity over the band, which excludes attenuation, omission and displacement (all with $E_p\le E_t$), and (b)~$\sigma_{\text{inc}}\approx\sigma_{\text{hf}}$, which excludes coherent amplification. Conversely, $E_p/E_t<1$ together with $\sigma_{\text{hf}}\le B$ and $\sigma_{\text{inc}}\ll\sigma_{\text{hf}}$ identifies attenuation or omission of the truth tail. Where the truth is empty above $\kappa_c$ (farfield patch), $T\approx0$ and all cases except additive content give $\sigma_{\text{hf}}\approx0$, so any residual there is content added by the model.

\begin{table}[htbp]
  \centering
  \caption{Response of the spectral diagnostics above $\kappa_c$ to controlled departures of the prediction from the truth, derived from their definitions. $T=\sum_{\kappa\ge\kappa_c}E_t\le B^2$ is the truth-tail energy and $E_n$ the energy of additive noise uncorrelated with the truth (expected values). For omission ($a=0$), $E_p=0$ and $\gamma^2$ is undefined.}
  \label{tab:controlled_cases}
  \resizebox{\textwidth}{!}{%
  \begin{tabular}{llcccc}
    \toprule
    Case & Prediction above $\kappa_c$ & $E_p/E_t$ & $\gamma^2$ & $\sigma_{\text{hf}}$ & $\sigma_{\text{inc}}$ \\
    \midrule
    Exact & $F_p=F_t$ & $1$ & $1$ & $0$ & $0$ \\
    Attenuation / omission & $F_p=aF_t$, $0\le a\le1$ & $a^2\le1$ & $1$ & $(1-a)\sqrt{T}\le B$ & $0$ \\
    Coherent amplification & $F_p=aF_t$, $a>1$ & $a^2>1$ & $1$ & $(a-1)\sqrt{T}$, $>B$ if $a>1+B/\sqrt{T}$ & $0$ \\
    Spatial displacement & $F_p=F_t\,e^{-2\pi\mathrm{i}\,\mathbf k\cdot\mathbf s/L}$ & $1$ & $\le1$ & $\le2\sqrt{T}$ & $\le\sqrt{T}$ \\
    Additive noise & $F_p=F_t+F_n$ & $1+E_n/E_t$ & $\approx E_t/(E_t+E_n)$ & $\approx(\sum E_n)^{1/2}$ & $\approx\sigma_{\text{hf}}$ \\
    \bottomrule
  \end{tabular}
  }
\end{table}

\begin{table}[htbp]
  \centering
  \caption{Probe-patch settings and derived scales of the spectral noise analysis for NACA8604 at $\alpha=-5^\circ$, state $v_y/U_\infty$ (lengths in chords $c$, wavenumbers in cycles/$c$). Cell sizes $h$ are Delaunay edge lengths of the cell centers inside the patch (minimum / median / 95th percentile).}
  \label{tab:psd_settings}
    \resizebox{\textwidth}{!}{%
      \begin{tabular}{lcc}
        \toprule
        & Farfield & Near-wake \\
        \midrule
        \multicolumn{3}{l}{\emph{Settings}} \\
        Patch center $(x_c,y_c)$ & $(0.50,\,2.00)$ & $(1.55,\,0.00)$ \\
        Patch extent & $x\in[0,1]$, $y\in[1.5,2.5]$ & $x\in[1.05,2.05]$, $y\in[-0.5,0.5]$ \\
        Side $L$ & $1$ & $1$ \\
        Grid $n_g\times n_g$ & $512\times512$ & $512\times512$ \\
        Grid spacing $\Delta=L/(n_g-1)$ & $1.96\times10^{-3}$ & $1.96\times10^{-3}$ \\
        Tail tolerance $\varepsilon$ & $10^{-3}$ & $10^{-3}$ \\
        Leakage floor $4/L$ & $4$ & $4$ \\
        \midrule
        \multicolumn{3}{l}{\emph{Derived scales}} \\
        Fundamental $\Delta\kappa=1/L$ & $1$ & $1$ \\
        Nyquist $\kappa_{\max}=(n_g-1)/(2L)$ & $256$ & $256$ \\
        Cell size $h$ (min / median / p95) & $0.0062$ / $0.037$ / $0.050$ & $0.0020$ / $0.0073$ / $0.015$ \\
        $\Delta\le h_{\min}/2$ & yes ($\Delta=0.32\,h_{\min}$) & marginal ($\Delta\approx h_{\min}$; median cell resolved $3.7\times$) \\
        Mesh Nyquist $\kappa_{\text{mesh}}=1/(2h)$ & $10$-$81$ & $34$-$250$ \\
        Interpolation imprint $\kappa\approx1/h$ & $20$-$160$ & $68$-$500$ \\
        Windowed truth variance $\operatorname{Var}_w(v_y)$ & $1.1\times10^{-5}$ & $2.4\times10^{-4}$ \\
        Cutoff $\kappa_c$ & $4$ & $18$ \\
        Binding constraint on $\kappa_c$ & leakage floor $4/L$ & truth tail (physics) \\
        Noise band $[\kappa_c,\kappa_{\max}]$ & $[4,256]$ & $[18,256]$ \\
        Content of $E_{p-t}$ above $\kappa_c$ & added content only (truth empty) & missing, amplified and/or added content \\
        Omission bound $B=(\varepsilon\operatorname{Var}_w v_y)^{1/2}$ & $1.0\times10^{-4}$ & $4.9\times10^{-4}$ \\
        \bottomrule
      \end{tabular}
    }
\end{table}

\paragraph{High-frequency residual.} Table~\ref{tab:noise_metric} reports $\sigma_{\text{hf}}=\sigma_\kappa(\kappa_c)$ for every model and state in both patches, together with its incoherent part $\sigma_{\text{inc}}=\big(\sum_{\kappa\ge\kappa_c}(1-\gamma^2)E_p\big)^{1/2}$ and the omission bound $B=(\varepsilon\operatorname{Var}_w q_t)^{1/2}$. Following the controlled cases, the columns are read together with the energy ratio: $\sigma_{\text{inc}}\approx\sigma_{\text{hf}}$ with $E_p/E_t>1$ means the residual above $\kappa_c$ is incoherent content the model adds (noise), whereas $\sigma_{\text{hf}}$ at or below $B$ with $\sigma_{\text{inc}}\ll\sigma_{\text{hf}}$ and $E_p/E_t<1$ means it is truth content the model attenuates or omits (over-smoothing). In the farfield patch the truth is empty, all entries are of the first kind ($\sigma_{\text{inc}}\approx\sigma_{\text{hf}}\gg B$), and MLP-NF is the noisiest model for every state. In the near-wake patch the velocity residuals of the NODE models lie at or below $B$ with a small incoherent part; together with $E_p/E_t\ll1$ above $\kappa_c$ (Figure~\ref{fig:spectral_noise} for $v_y$), this identifies attenuation of the shear-layer scales rather than added noise. MLP-NF instead exceeds $B$ with a mostly incoherent residual, i.e. added noise. The pressure residual of the NODE models also exceeds its (very small) bound with $\sigma_{\text{inc}}\approx\sigma_{\text{hf}}$; since $\sigma_{\text{hf}}>2B$, it can be explained neither by displacement nor by attenuation of the truth tail and is noise-like, although at an absolute level below $10^{-4}\,p_\infty$.

\begin{table}[htbp]
  \centering
  \caption{High-frequency residual $\sigma_{\text{hf}}$ and its incoherent part $\sigma_{\text{inc}}$ for NACA8604 at $\alpha=-5^\circ$ per model, state and patch (velocities in units of $U_\infty$, pressure in units of $p_\infty$). Cutoffs from the truth-tail criterion: farfield $\kappa_c=(5,\,4,\,4)$ and near-wake $\kappa_c=(29,\,18,\,5)$ cycles/$c$ for $(v_x,\,v_y,\,p)$. The last row of each block is the omission bound $B=(\varepsilon\operatorname{Var}_w q_t)^{1/2}$, the level reached by a prediction that omits the truth tail (Appendix~\ref{sec:Spectral Noise analysis}); it is not an upper bound for every noise-free prediction.}
  \label{tab:noise_metric}
  \resizebox{\textwidth}{!}{%
  \begin{tabular}{llcccccc}
    \toprule
    & & \multicolumn{3}{c}{$\sigma_{\text{hf}}$} & \multicolumn{3}{c}{$\sigma_{\text{inc}}$} \\
    \cmidrule(lr){3-5}\cmidrule(lr){6-8}
    Patch & Model & $v_x$ & $v_y$ & $p$ & $v_x$ & $v_y$ & $p$ \\
    \midrule
    Farfield  & MLP-NF   & $3.5\times10^{-3}$ & $4.7\times10^{-3}$ & $2.4\times10^{-4}$ & $3.5\times10^{-3}$ & $4.7\times10^{-3}$ & $2.3\times10^{-4}$ \\
              & MLP-NODE & $2.0\times10^{-3}$ & $1.8\times10^{-3}$ & $1.6\times10^{-4}$ & $1.9\times10^{-3}$ & $1.7\times10^{-3}$ & $1.5\times10^{-4}$ \\
              & GNN-NODE & $1.7\times10^{-3}$ & $2.5\times10^{-3}$ & $1.3\times10^{-4}$ & $1.6\times10^{-3}$ & $2.4\times10^{-3}$ & $1.3\times10^{-4}$ \\
              & omission bound $B$ & $1.1\times10^{-4}$ & $1.0\times10^{-4}$ & $5.6\times10^{-6}$ & -- & -- & -- \\
    \midrule
    Near-wake & MLP-NF   & $4.0\times10^{-3}$ & $3.9\times10^{-3}$ & $5.6\times10^{-4}$ & $2.9\times10^{-3}$ & $3.9\times10^{-3}$ & $5.5\times10^{-4}$ \\
              & MLP-NODE & $2.6\times10^{-3}$ & $4.4\times10^{-4}$ & $8.9\times10^{-5}$ & $3.2\times10^{-4}$ & $4.7\times10^{-5}$ & $8.5\times10^{-5}$ \\
              & GNN-NODE & $2.8\times10^{-3}$ & $4.3\times10^{-4}$ & $8.4\times10^{-5}$ & $1.8\times10^{-4}$ & $3.5\times10^{-5}$ & $8.3\times10^{-5}$ \\
              & omission bound $B$ & $2.8\times10^{-3}$ & $4.9\times10^{-4}$ & $2.6\times10^{-5}$ & -- & -- & -- \\
    \bottomrule
  \end{tabular}
  }
\end{table}

\section{Generalization to Non-NACA Airfoils}\label{app:non-NACA airfoils}
This appendix complements the non-NACA airfoil generalization study with the full per-AoA results for the five non-NACA airfoils. The profiles, beyond their characteristic application to automotive, were chosen also because they span a range of camber and thickness well outside the NACA 4-digit training set. NASA LS(1)-0413 is used in Grand Touring rear wings for its high camber and stall-resistant behavior near separation; Selig S1223 is a very high-lift, highly cambered profile common in low-to-moderate Reynolds number and Formula Student front-wing studies; Eppler E423 is a classical high-lift benchmark used in automotive rear-wing and Formula-1-inspired spoiler studies; Clark Y is a simple, widely used reference profile for inverted racing car wings; and Wortmann FX 63-137 offers efficient lift and strong lift-to-drag performance at modest Reynolds numbers.

Table~\ref{tab:l2_non_naca} reports the $\mathcal{L}_2$-norm error of $v_x/U_\infty$ over the full domain for each airfoil and model across the AoA range. Three observations stand out. First, the error scales with how far the geometry departs from the training distribution: the mildly cambered Clark Y is predicted most accurately (mean error $0.67$--$0.84\%$), followed by NASA LS(1)-0413 ($1.37$--$1.42\%$), whereas the highly cambered Eppler E423, Wortmann FX 63-137 and, in particular, Selig S1223 ($2.64$--$2.79\%$) yield the largest errors. Second, MLP-NODE is the most accurate model in the majority of cases, most consistently on Clark Y and Eppler E423 where it is the best model at every AoA; MLP-NF is competitive on Wortmann FX 63-137 and NASA LS(1)-0413, and GNN-NODE is rarely the best but is never far behind. Third, the error grows with $|\alpha|$ for all airfoils, with the strongest sensitivity for the high-camber profiles S1223 and FX 63-137 (CoV of $15$--$32\%$), followed by E423 ($8$--$14\%$), versus $6$--$9\%$ for Clark Y and LS(1)-0413. The clearest example is Selig S1223, where MLP-NODE degrades fastest at large $|\alpha|$ (reaching $4.28\%$ at $\alpha=-10^\circ$, CoV of $32.2\%$), so that MLP-NF, although less accurate at small $|\alpha|$, becomes the most robust model beyond $\alpha=-8^\circ$. Even in this worst case the error remains below $4.3\%$, which supports the expectation that the models would continue to generalize well if the training set were expanded with additional airfoil families.

\begin{table}[htbp!]
  \centering
  \caption{$\mathcal{L}_2$-norm based error (\%) of $v_x/U_\infty$ over the full computational domain for models MLP-NF, MLP-NODE, GNN-NODE. Five non-NACA airfoils listed: Selig S1223, Clark Y, Eppler E423, Wortmann FX 63-137 and NASA LS(1)-0413.}
  \label{tab:l2_non_naca}
  \resizebox{\textwidth}{!}{%
  \begin{tabular}{c c c c c c c c c c c c c c c c}
    \toprule
    & \multicolumn{3}{c}{Selig S1223} & \multicolumn{3}{c}{Clark Y} & \multicolumn{3}{c}{Eppler E423} & \multicolumn{3}{c}{Wortmann FX 63-137} & \multicolumn{3}{c}{NASA LS(1)-0413} \\
    \cmidrule(lr){2-4} \cmidrule(lr){5-7} \cmidrule(lr){8-10} \cmidrule(lr){11-13} \cmidrule(lr){14-16}
    $\alpha$ $[^\circ]$ & NF & MLP & GNN & NF & MLP & GNN & NF & MLP & GNN & NF & MLP & GNN & NF & MLP & GNN \\
    \midrule
    0 & 2.09 & \textcolor{green}{1.81} & 2.08 & 0.76 & \textcolor{green}{0.62} & 0.69 & 1.79 & \textcolor{green}{1.53} & 2.00 & \textcolor{green}{1.41} & 1.42 & 1.67 & 1.22 & \textcolor{green}{1.21} & 1.27 \\
    -1 & 2.17 & \textcolor{green}{1.86} & 2.11 & 0.75 & \textcolor{green}{0.60} & 0.65 & 1.82 & \textcolor{green}{1.53} & 1.99 & \textcolor{green}{1.46} & 1.47 & 1.72 & \textcolor{green}{1.26} & 1.28 & 1.32 \\
    -2 & 2.26 & \textcolor{green}{1.92} & 2.16 & 0.76 & \textcolor{green}{0.61} & 0.65 & 1.87 & \textcolor{green}{1.56} & 2.00 & \textcolor{green}{1.52} & 1.53 & 1.78 & \textcolor{green}{1.30} & 1.34 & 1.37 \\
    -3 & 2.36 & \textcolor{green}{2.01} & 2.23 & 0.78 & \textcolor{green}{0.61} & 0.67 & 1.94 & \textcolor{green}{1.61} & 2.04 & \textcolor{green}{1.60} & 1.61 & 1.84 & \textcolor{green}{1.32} & 1.37 & 1.39 \\
    -4 & 2.51 & \textcolor{green}{2.15} & 2.37 & 0.80 & \textcolor{green}{0.62} & 0.69 & 2.04 & \textcolor{green}{1.70} & 2.11 & \textcolor{green}{1.69} & 1.70 & 1.93 & \textcolor{green}{1.35} & 1.41 & 1.41 \\
    -5 & 2.66 & \textcolor{green}{2.32} & 2.56 & 0.84 & \textcolor{green}{0.65} & 0.72 & 2.15 & \textcolor{green}{1.79} & 2.20 & \textcolor{green}{1.80} & \textcolor{green}{1.80} & 2.04 & \textcolor{green}{1.38} & 1.44 & 1.43 \\
    -6 & 2.85 & \textcolor{green}{2.59} & 2.82 & 0.87 & \textcolor{green}{0.68} & 0.76 & 2.26 & \textcolor{green}{1.86} & 2.30 & 1.91 & \textcolor{green}{1.88} & 2.15 & \textcolor{green}{1.36} & 1.41 & 1.41 \\
    -7 & 3.04 & \textcolor{green}{2.94} & 3.13 & 0.88 & \textcolor{green}{0.70} & 0.78 & 2.36 & \textcolor{green}{1.91} & 2.37 & 2.04 & \textcolor{green}{1.99} & 2.29 & \textcolor{green}{1.38} & 1.42 & 1.43 \\
    -8 & \textcolor{green}{3.24} & 3.37 & 3.47 & 0.90 & \textcolor{green}{0.72} & 0.79 & 2.46 & \textcolor{green}{1.97} & 2.43 & 2.18 & \textcolor{green}{2.11} & 2.41 & \textcolor{green}{1.43} & 1.46 & 1.48 \\
    -9 & \textcolor{green}{3.43} & 3.82 & 3.77 & 0.92 & \textcolor{green}{0.74} & 0.80 & 2.54 & \textcolor{green}{2.02} & 2.44 & 2.30 & \textcolor{green}{2.24} & 2.50 & \textcolor{green}{1.49} & 1.52 & 1.52 \\
    -10 & \textcolor{green}{3.63} & 4.28 & 4.03 & 0.95 & \textcolor{green}{0.78} & 0.83 & 2.59 & \textcolor{green}{2.02} & 2.36 & 2.36 & \textcolor{green}{2.33} & 2.50 & 1.58 & 1.60 & \textcolor{green}{1.57} \\
    \midrule
    mean & 2.75 & \textcolor{green}{2.64} & 2.79 & 0.84 & \textcolor{green}{0.67} & 0.73 & 2.17 & \textcolor{green}{1.77} & 2.20 & 1.84 & \textcolor{green}{1.83} & 2.08 & \textcolor{green}{1.37} & 1.40 & 1.42 \\
    std & 0.53 & 0.85 & 0.71 & 0.07 & 0.06 & 0.06 & 0.30 & 0.20 & 0.18 & 0.34 & 0.31 & 0.31 & 0.10 & 0.11 & 0.09 \\
    min & 2.09 & 1.81 & 2.08 & 0.75 & 0.60 & 0.65 & 1.79 & 1.53 & 1.99 & 1.41 & 1.42 & 1.67 & 1.22 & 1.21 & 1.27 \\
    max & 3.63 & 4.28 & 4.03 & 0.95 & 0.78 & 0.83 & 2.59 & 2.02 & 2.44 & 2.36 & 2.33 & 2.50 & 1.58 & 1.60 & 1.57 \\
    CoV\,(\%) & 19.2 & 32.2 & 25.2 & 8.4 & 9.1 & 8.7 & 13.7 & 11.1 & 8.3 & 18.4 & 17.2 & 15.0 & 7.4 & 7.6 & 6.1 \\
    \bottomrule
  \end{tabular}
  }
\end{table}

\begin{figure}[hbpt!]
    \centering
    \includegraphics[width=0.8\linewidth]{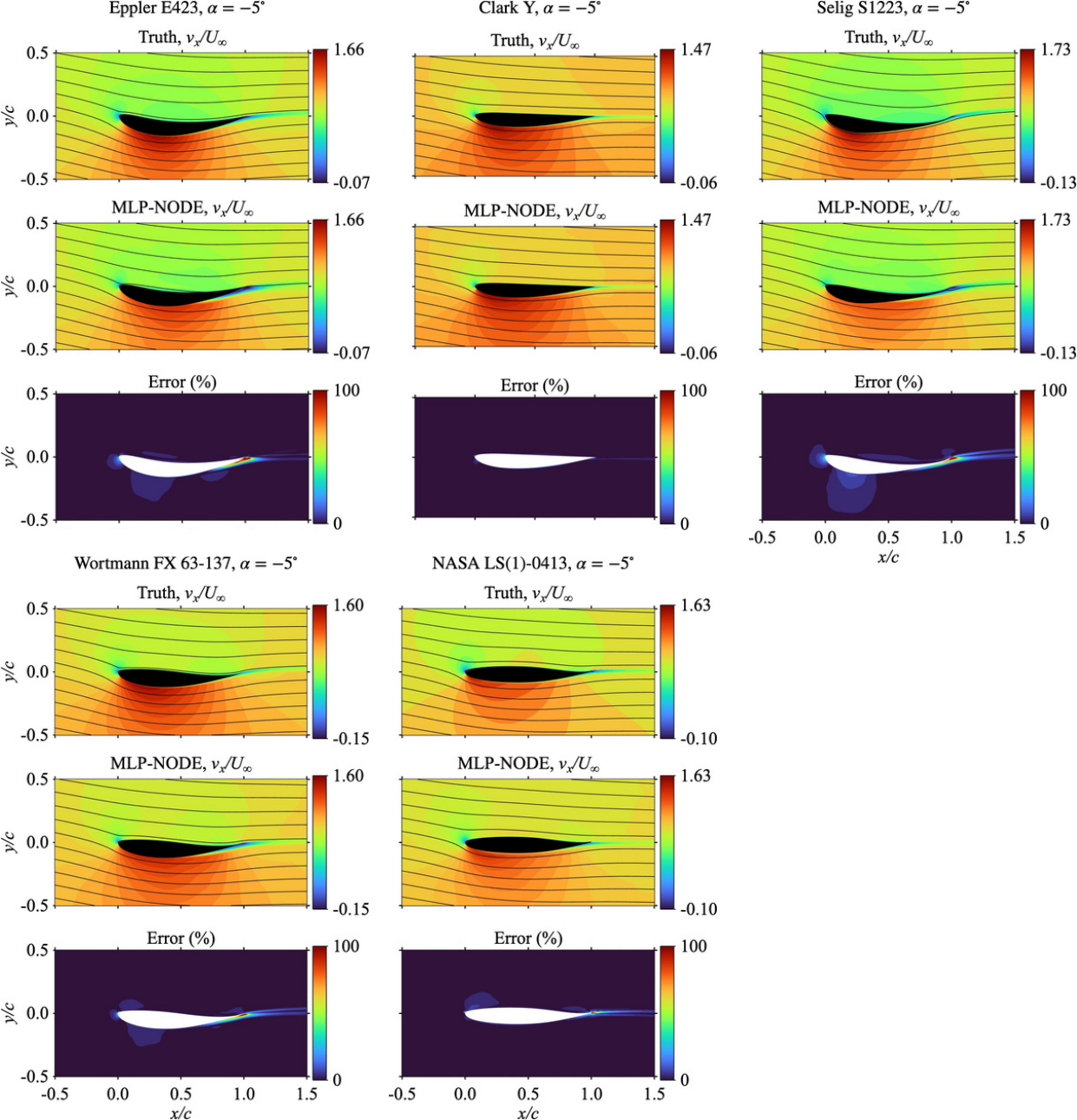}
    \caption{Comparison of $v_x/U_\infty$ at $\alpha=-5^{\circ}$ between OpenFOAM and the MLP-NODE prediction. This case illustrates the model’s ability to extrapolate airfoil geometries beyond NACA 4-digit airfoils using only NACA training data. Five non-NACA airfoils are represented: Eppler E423, Clark Y, Selig S1223, Wortmann FX 63-137 and NASA LS(1)-0413.}
    \label{fig:non_naca_mlp}
\end{figure}

\begin{figure}[hbpt!]
    \centering
    \includegraphics[width=0.7\linewidth]{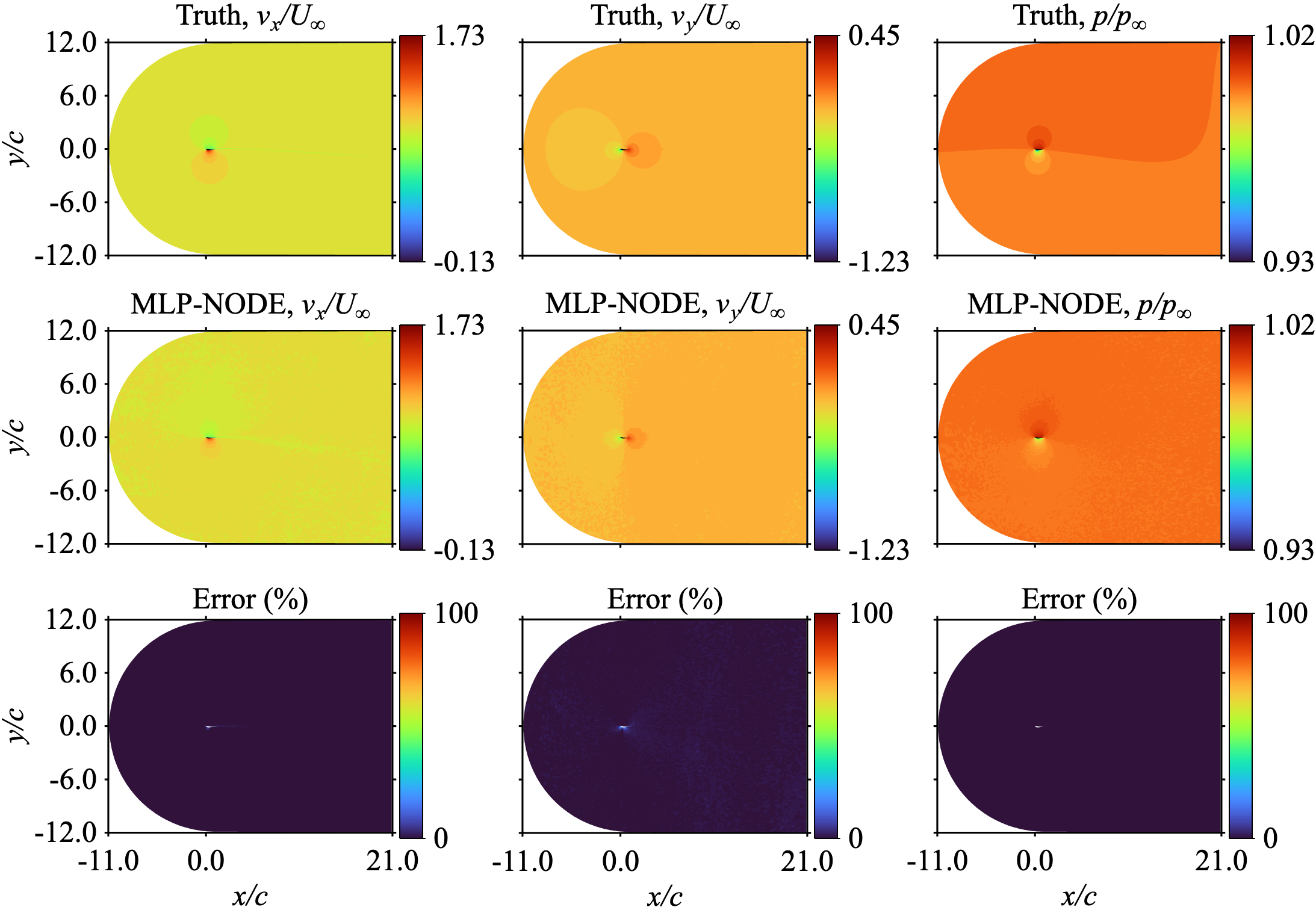}
    \caption{Full flow field comparison of (first column) $v_x/U_\infty$, (second column) $v_y/U_\infty$ and (third column) $p/p_\infty$ between OpenFOAM and the MLP-NODE prediction for Selig S1223 airfoil at $\alpha = -5^{\circ}$.}
    \label{fig:full_field}
\end{figure}

Figures~\ref{fig:non_naca_mlp} and~\ref{fig:full_field} show the corresponding predicted fields for the MLP-NODE model at $\alpha=-5^\circ$. Figure~\ref{fig:non_naca_mlp} plots the contours of streamwise velocity $v_x$ and streamlines of the five non-NACA airfoils in a field of view around the airfoil. The point-wise percentage error is remarkably low everywhere in the flow except for few regions notoriously well-known for challenging predictive models also highlighted and discussed in the NACA 4-digit series field reconstruction: the wake, the stagnation point and the boundary layer. The most inaccurate region is at the onset of the wake development whose precursor is flow separation in the lower airfoil surface. The most cambered airfoils in this restricted set (S1223, E423 and FX 63-137) display the most inaccurate wake reconstructions, while Clark Y and NASA LS(1)-0413 flow fields are exceptionally well predicted by the MLP-NODE model.

Finally, Figure~\ref{fig:full_field} depicts the full flow field view of Selig S1223 airfoil at $\alpha=-5^\circ$. The MLP-NODE model retrieves the constant farfield values of the velocity and pressure states, with additional small-scale, random noise scattered over the computational domain. This low-level noise present in regions of the flow with little or no physical significance, quantified in Figure~\ref{fig:spectral_noise} and Table~\ref{tab:noise_metric} for NACA8604, is invisible in the point-wise percentage error maps, but adds up in the overall $\mathcal{L}_2$-norm error quantification. Therefore, restricting the domain for model training may result in a more optimal usage of computational resources towards model optimization and enhanced predictive accuracy without loss of physics.

\end{document}